\documentstyle[psfig,aps,tighten,eqsecnum,twocolumn]{revtex}

\begin{document}

%%%%%%%%%%%%%%%%% MACROS %%%%%%%%%%%%%%%%%%%%%%%%%%

\newcommand{\vvs}{\vspace{0.5cm}}
\newcommand{\vs}{\vspace{0.1cm}}

\newcommand{\be}{\begin{equation}}
\newcommand{\ee}{\end{equation}}
\newcommand{\ba}{\begin{eqnarray}}
\newcommand{\ea}{\end{eqnarray}}

\newcommand{\fixme}[1]{{\bf [FIXME: #1]}}

\newcommand{\feyn}[1]{\rlap{\,/}{#1}}
\newcommand{\tr}{\mathop{\rm tr}\nolimits}
\hyphenation{analysis}

%%%%%%%%%%%%%%%%%%%%% TITLE %%%%%%%%%%%%%%%%%%%%%%%%%%%%%%%%

\title{Hadronization of a Quark-Gluon Plasma in the Chromodielectric Model}

\author{
 {\sc C.~T.~Traxler$^1$, U.~Mosel} \\
 {\small Institut f\"ur Theoretische Physik, Universit\"at Giessen} \\
 {\small D-35392 Giessen, Germany} \\
 and \\
 {\sc T.~S.~Bir\'o} \\
 {\small Research Institute for Particle and Nuclear Physics, Budapest} \\
 {\small H-1525 Budapest, Hungary}
}

\date{\today}

\maketitle
\footnotetext[1]{part of the Ph.~D. thesis of C.~T.~Traxler}

%%%%%%%%%%%%%%%%%%%%%%%%%%% ABSTRACT %%%%%%%%%%%%%%%%%%%%%%%%%%%

\begin{abstract}
  We have carried out simulations of the hadronization of a hot, ideal
  but effectively massive quark-gluon gas into color neutral clusters
  in the framework of the semi-classical SU(3) chromodielectric
  model.
  We have studied the possible quark-gluon compositions of clusters
  as well as the final mass distribution and spectra,
  aiming to obtain an insight into relations between hadronic
  spectral properties and the confinement mechanism in this model.
  \centerline{{\bf pacs: {12.38.Mh, 12.38.Lg, 24.85+p}} }
\end{abstract}

%%%%%%%%%%%%%%%%%%%%%%%%%%%% INTRODUCTION %%%%%%%%%%%%%%%%%%%%

\vvs
\section{Introduction}

\vs
The quest for the quark-gluon plasma is one of the major scientific
efforts of the last decade of strong interaction physics.
In recent years $200$ A\,GeV $Pb$ beams became available for the
experimental study at CERN SPS, and future colliders will reach even
higher energies, namely $200 + 200$ A\,GeV at RHIC Brookhaven and 
$900 + 900$ A\,GeV at CERN LHC.
These energies correspond to an estimated maximal energy density
of \hbox{$\varepsilon = 5$ GeV/fm$^3$}  and
\hbox{$\varepsilon = 9$ GeV/fm$^3$} or temperature 
\hbox{$T = 230$ MeV} and \hbox{$T = 260$ MeV,} respectively.
At CERN SPS the achieved maximal temperature is about 
\hbox{$T = 160$ MeV}\cite{QM91}.

Heavy ions like $Au$ and $Pb$ are the best means for approaching
the thermodynamical limit in particle accelerator experiments
as closely as possible. They provide high energy- and baryon
densities.
These advantages on the experimental side lead to trouble for
theoretical studies: although the thermodynamical approach to 
quark matter is highly developed\cite{lattice,high-T}, 
a dynamical insight or finite baryon density studies are at their 
very beginning\cite{Keldish,Bielefeld}.
The solution of both theoretical problems, i.e. the
real-time description and the study at finite density,
is connected to the invention of new non-perturbative techniques,
which are able to handle complex effective actions.
On the other hand,
there are several phenomenological models aiming to simulate
these unknown dynamical features of quantum chromodynamics (QCD).

In theoretical works the main goal is to understand the
mechanism of {\em color confinement}. In this perspective, 
understanding hadronization means learning about the confinement
mechanism in an experimentally controllable situation.
We briefly review hadronization models presently known to us
grouped in three categories: {\em prompt,}
{\em equilibrium} and {\em evolutionary.}
The process takes zero (i.e. less than 1 fm$/c$),
infinite (more than 20 fm$/c$)
or a finite amount of time (2 - 6 fm$/c$) in these three
respective categories.

String- and parton-fragmentation models
\cite{LUND,PAR-CAS} in the first category
are the oldest prompt 
hadronization models; they have been 
used already in the description of elementary
particle collisions ($pp$, $p\bar p$ and $e^+e^-$).
With the help of Glauber's model \cite{GLAUB} the yields of
elementary $pp$-events are then summed up for a complicated
heavy ion collision. In these calculations hadron ratios
depend on phenomenological fragmentation ratios or
fragmentation functions \cite{FEYNMAN}. Although these parameters
are non-perturbative, any explanation or derivation of them
from QCD is missing.

Further statistical and combinatorical models connect quark numbers
and (pre-)hadron numbers by assuming an underlying coalescence.
The quark cluster model used e.g. in the parton cascade
model \cite{GEIGER-MULLER} or partially in string models like
VENUS \cite{WERNER} and (U)RQMD \cite{RQMD}
is sensitive to quark properties, like
effective masses. The ALCOR model \cite{ZIMANYI} assumes an extremely 
fast and far-from-equilibrium hadronization of massive quarks 
after gluons are fragmented. Although individual
hadron formation is immediate, the formation time
instant may vary over a range of several tens of fm$/c$ in
these approaches.
In these models the hadron formation
is a process, whose explanation is still absent.

In contrast, the mixed phase model \cite{TONEEV} 
--- one of the most recent representatives of the
second category ---
focuses on the state of thermodynamical equilibrium. 
It constructs a non-ideal equation of state
which includes part of the particle correlations and is able
to show confinement behavior.
Although --- in our opinion --- this is a promising research direction,
it fails to describe a very non-equilibrium hadronization, where
the characteristic time of bound quark cluster formation is commensurate
to the thermal equilibrium maintenance time.
Also, diquarks and baryons have not been included yet.

There are several chemical approaches to hadronization,
showing an evolution time of a few to several fm$/c$.
Assuming a massive, effectively non-relativistic quark matter,
the hadronization can be described by utilizing phenomenological
confinement potentials or just the perturbative Coulomb-like
potential. The elimination of color charge during this
process is then due to medium effects, like string formation.
The time evolution of the mixed quark-hadron system is
described by a set of rate equations. Such an approach
has been published a decade ago \cite{KNOLL+BARZ}, and is
recently under development in the framework of ALCOR model\cite{PRIVATE}.
 
Also in the framework of Nambu Jona-Lasinio (NJL) model 
hadronization cross sections have been obtained \cite{KLEVANSKY}.
Since mesons, like pions and $\sigma$-mesons, are elementary
in this model, the confinement mechanism,
an important detail of the microscopic process, remains
unexplored. An alternative model, the chromodielectric model (CDM)
originally proposed by Friedberg and Lee \cite{LEE},
describes hadrons as clusters of quarks bound in a phenomenological
scalar potential. While there have been a tantamount number of
studies of stationary states of this model with quantum-mechanical
quarks \cite{WILETS,RIPKA,MOSEL,KALMBACH}, a real-time dynamical approach
is quite recent \cite{VETTER,PHDVETTER,LOH,PHDLOH}. 
The latter studies utilized semi-classical
transport theory, treating quarks like classical charge clouds.

In the present paper we report about a classical molecular dynamics
approach on the quark level to hadronization. For the dynamical 
description, the classical limit of the chromodielectric model
is used, where quarks and gluons are represented by extended but rigid
charge distributions. The background fields are a scalar field
and two abelian fields, corresponding to the two neutral colors
in $SU(3)$. The six remaining, charged, gluons are treated as particles.
We deal with these simplifications in
order to be able to carry out a molecular dynamical calculation
of about 400 quarks in an acceptable time. 
The essential confinement mechanism in this model is hidden in
the scalar field self-interaction potential and the construction of
the chromodielectric constant, whose particular
dependence on the effective scalar field $\sigma$ includes
the main quantum-mechanical effect from QCD.

We think that such microscopical studies may reveal whether
or not the hadronization process is faster than the maintenance
of thermal or chemical
equilibrium, whether or not the formed hadrons are close to the
predictions of constituent quark models, and what kind of
color-neutral clusters form in abundance.
We think that the drawbacks of this approach, namely the classical treatment
of the color charge and its clusters, the abelian dominance
assumption in the long-range fields, and the use of the
$\sigma$-field, are outweighted by the advantages: an event by
event, microscopic, dynamical simulation including flux-tubes that are
non-elementary but formed by a dynamical process.
This is useful for studies of possible hadronization scenarios of
quark matter until real-time QCD calculations become feasible.

This paper is organized as follows: 
in Sec.~\ref{ChromodielectricModel}, a brief
description of the original chromodielectric model is followed
by an explanation of the confinement mechanism in this model.
Section \ref{TowardsAComputerSimulation} presents our
practical approach to the model: we deal with the treatment of 
color charges in the classical limit, present our notion of hadrons 
within the model, and touch upon relevant computational
methods. In particular, the principles of a fast adaptive multigrid
finite element procedure for the solution of the Gauss law are discussed.
Section \ref{StaticalModelProperties} reports on our choice of model
parameters and its consequences on statical model properties.
Section \ref{HadronizationOfQGP} is the main part of the present
article: it presents actual simulation runs of hadronization 
scenarios. We describe our choice of initial states and show snapshots of
the program runs. We then compile our results on cluster-formation, 
and show mass, rapidity, and transverse momentum distributions for white
clusters. Finally, Sec.~\ref{Conclusion} summarizes our conclusions.

%%%%%%%%%%%%%%%%%%%%%%%  MODEL %%%%%%%%%%%%%%%%%%%%%%%%%%%%%

\vvs
\section{The Chromodielectric Model}
\label{ChromodielectricModel}

\vs
Since free quarks have never been observed experimentally,
it is generally accepted that any physical state of finite energy has
to be a singlet under color-$SU(3)$ transformations. This phenomenon 
is called color confinement. Since we know that classical (``tree-level'') 
chromodynamics does not enforce color confinement,  
this has to be a quantum (``loop'') effect, characteristic of a
strongly interacting non-abelian gauge field theory.
However, a proof of color confinement based solely on the QCD Lagrangian
has not yet been given, mainly because it is too difficult to identify the
relevant QCD degrees of freedom.

The chromodielectric (Friedberg-Lee) model \cite{LEE} tries to mimic QCD's
color confinement {\em on the classical level.} It introduces an
additional scalar field $\sigma$ that enforces color confinement in a 
conceivable way. 

%%%%%%%%%%%%%%%%%%%%%%%  MODEL LAGRANGIAN %%%%%%%%%%%%%%%%%%%%%%%%%%%%%

\vvs
\subsection{Model Lagrangian}

\vs
The Lagrangian reads
\be
{\cal L}\ =\ {\cal L}_p+{\cal L}_c+{\cal L}_\sigma
\label{Lagrangian}
\ee 
with
\ba
{\cal L}_p&=& \sum_{f}\bar q^f\left(i\feyn D - m_f \right)q^f
\label{ParticleLagrangian}\\
{\cal L}_c&=& -{\kappa(\sigma)\over 4} F_{\mu\nu}^a F^{\mu\nu a}
\label{ColorLagrangian}\\
{\cal L}_\sigma&=& {1\over 2}(\partial_\mu \sigma)(\partial^\mu\sigma)-U(\sigma).
\label{SigmaLagrangian}
\ea 
The symbol $q^f$ denotes a quark spinor of flavor $f$; both the color and
Dirac indices are suppressed, and the Feynman dagger notation 
\mbox{$\feyn D=\gamma^\mu D_\mu$} is used.
The covariant derivative \mbox{$iD_\mu=i\partial_\mu-g_v A_\mu$} contains the
color vector potential \mbox{$A_\mu=A_\mu^a{\lambda^a\over2}$}, where
the $\lambda^a$ denote the eight Gell-Mann matrices. 
The color field tensor is defined as 
\be
F_{\mu\nu}\ =\ F_{\mu\nu}^a{\lambda^a\over2}
   \ :=\ {i\over g_v}[D_\mu, D_\nu].
\label{FieldtensorDefinition}
\ee
It obeys the classical chromodynamical field equation
in a covariant medium with dielectric function $\kappa(\sigma)$,
\be
  [D_\mu, \kappa(\sigma)F^{\mu\nu}]\ =\ j^\nu,
\label{YangMills}
\ee
with color current 
\ba 
 j^\nu&=& g_v\sum\limits_f \bar q^f \gamma^\nu \lambda^a 
          q^f{\lambda^a\over 2} \nonumber \\
      &=:& j^{\nu a}{\lambda^a\over 2}.
\label{ColorCurrent}
\ea

The quark field obeys a Dirac equation,
\be
 \left(i\feyn D - m_f\right)q^f\ =\ 0
\label{DiracEquation}
\ee
with a constituent quark mass $m_f$.
The quarks can in principle be treated quantum-mechanically here, 
while the color and $\sigma$ field are regarded as classical fields.

\begin{figure}
\centerline{\psfig{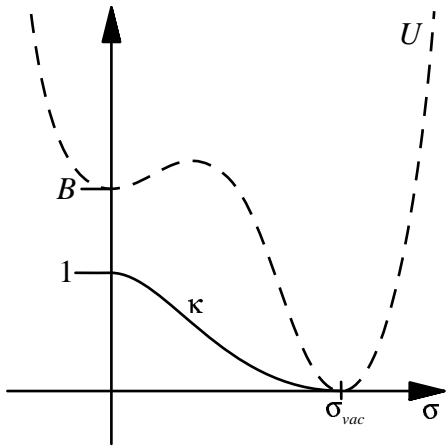}}
\caption{\em Schematical graph of the functions $U(\sigma)$ and
$\kappa(\sigma)$. Note that $U$ is an energy density while $\kappa$ is
dimensionless. $\sigma_{vac}$ denotes the value of the $\sigma$ field
in the physical vacuum (ground state).}
\label{Ukappa}
\end{figure}

The scalar field $\sigma$ obeys a Klein-Gordon equation 
with a fourth-order polynomial self-interaction $U(\sigma)$ and a
complicated source term 
\be
  \partial_\mu\partial^\mu\sigma+U'(\sigma)
 \ =\ -{\kappa'(\sigma)\over4}F_{\mu\nu}^a F^{\mu\nu a}
\label{sigmaFieldEquation1}
\ee
Both functions $U$ and $\kappa$ are drawn schematically in Fig.~\ref{Ukappa}. 
Note that there is no direct coupling between the quarks and the
$\sigma-$field in this model.

%%%%%%%%%%%%%%%%%%%%%%%  CONFINEMENT MECHANISM %%%%%%%%%%%%%%%%%%%%%%%%%%%%%

\vvs
\subsection{Confinement Mechanism}
\label{ConfinementMechanism}

\vs
The Lagrangian (\ref{Lagrangian}) describes a classical field
theory with an explicit color confinement mechanism.
The ingredients of this mechanism are the self-interaction term
$U(\sigma)$, the dielectric function $\kappa(\sigma)$, and the color 
Gauss law. In this section, we utilize a simple example to 
explain how these three ingredients work together to enforce color 
confinement.

First note that in the vacuum of this theory, \mbox{$\sigma\equiv\sigma_{vac}$}
holds everywhere and thus \mbox{$\kappa\equiv0$}.
\begin{figure}
\centerline{\psfig{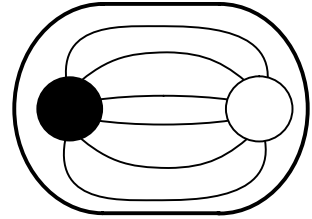}}
\caption{\em A color flux tube.}
\label{string}
\end{figure}
Now consider the static and color neutral quark-antiquark configuration depicted in
Fig.~\ref{string}.
The charge distribution is given by 
\mbox{$j_0(x) = j^3_0(x)\frac{\lambda^3}{2}$}, i.~e. it belongs to a
$U(1)$ subgroup of $SU(3)$. The Gauss law reads for this case
\be
  \vec \nabla \cdot \vec D^3\ =\ j^3_0.
\label{GaussLaw3}
\ee
It forces a color flux 
\mbox{$\vec D^3=\kappa(\sigma)\left(-\vec\nabla A_0^3-\partial_t \vec A^3\right)$} 
to stretch from the quark to the antiquark. 
This flux cannot exist in regions where
\mbox{$\kappa=0$}, therefore the $\sigma$ field has to deviate 
from its vacuum value,
forming a tube (or bag) with \mbox{$\sigma<\sigma_{vac}$} for the flux. 
Assuming $U$ and $\kappa$ are defined so that $\sigma$ ends up being
zero in the bag, the bag formation process costs an energy density 
$B$, which in turn can be interpreted as the vacuum pressure exerted
on the bag. 

Formulated differently, the vacuum pressure generated by
$U(\sigma)$ and the coupling term $\kappa F^2$ confines the color 
flux to a string --- {\em $U$ and $\kappa$ together generate the color flux 
confinement}. On the other hand, the quarks cannot escape from this string
because {\em the Gauss law demands a contiguous color flux to connect them.}
Unless pair creation is allowed for, the resulting string can become 
arbitrarily long, causing the $q\bar q-$potential to rise linearly. 

Let us contrast this with the corresponding result in classical chromodynamics.
The Gauss law, reduced to the abelian subgroup of our problem, reads then
\be
  \vec \nabla \cdot \vec E^3\ =\ j^3_0
\label{GaussLaw3p}
\ee
The solution is --- like in electrodynamics --- 
simply the linear superposition of the Coulomb
fields of the two charges. Since the long-distance Coulomb 
potential is $\propto 1/r$ and not linear, we see that 
classical chromodynamics does indeed not include color confinement. 

%%%%%%%%%%%%%%%%%%%%%%%  TREATMENT OF COLOR %%%%%%%%%%%%%%%%%%%%%%%%%%%

\vvs
\section{Treatment of Color Charges, Field Equations, and Hadrons}
\label{TowardsAComputerSimulation}

\vvs
\subsection{A Classical Molecular Dynamics Approach for the Charges}
\label{MolecularDynamicsApproach}

\vs
On our way to build a computer simulation of the chromodielectric model,
we have to make several approximations. In doing so, we take care
to preserve color confinement, since this is what we want to simulate. 

We treat only the two commuting gluon fields $A_\mu^3\frac{\lambda^3}{2}$ 
and $A_\mu^8\frac{\lambda^8}{2}$ as classical fields; they 
correspond to two decoupled, color neutral fields which behave just
like Maxwell fields. So from now on, the color index $a$ is generally 
restricted to the values 3 and 8. This approach amounts to taking a 
$U(1)\times U(1)$ abelian subgroup of the full $SU(3)$ gauge group of
QCD, and is therefore called the {\em abelian approximation}\cite{ELZE-HEINZ,LOH}. 
However, in contrast to the cited works, we do not simply drop the
other, charged, gluons. Instead, they are treated like the
quarks and antiquarks: as classical charge distributions with
a Gaussian shape of fixed width. Spin and isospin are accounted for 
by appropriate degeneracy factors.
This changes the particle Lagrangian
(\ref{ParticleLagrangian}) into the classical expression
\be
{\cal L}^{class}_p
\ =\ - \sum_i \rho_N(\vec x-\vec x_i) m_i \sqrt{1-\dot{\vec x}_i^2}\,
-j^{\mu a}A_\mu^a
\label{ClassicalParticleLagrangian}
\ee
with the color current
\be
j^{\mu a}\ =\ g_v\sum_i \rho_N(\vec x-\vec x_i)q_i^a {1\choose \dot{\vec x}_i}.
\label{ClassicalColorCurrent}
\ee
The index $i$ runs over all quarks,
antiquarks and charged gluons in the system.

\begin{figure}
\centerline{\psfig{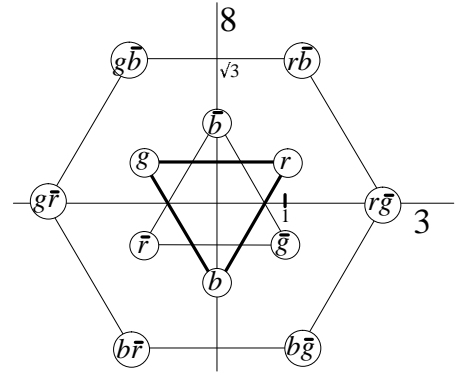}}
\caption{\em Quark and gluon colors.
  The bold equilateral triangle corresponds to the quark colors,
  the other triangle to the antiquark colors, and the regular hexagon
  to the six charged gluons. The denotations $r$, $g$, $b$, $\bar r$,
  $\bar g$, $\bar b$ are chosen arbitrarily.}
\label{colors}
\end{figure}

For the quarks, the color charges $q_i^{3,8}$ in
(\ref{ClassicalColorCurrent}) are equal to
the diagonal entries in $\lambda^{3,8}$.
Antiquarks carry the charge $-q_i^{3,8}$, and
the gluons carry each the sum of one quark and one antiquark
color charge. Table~\ref{ColorTable} and Fig.~\ref{colors} 
summarize the possible values.

The masses $m_i$ are model parameters; our particular choices
for them are listed and explained in Sec.~\ref{ModelParameters}.
The function $\rho_N$ is a Gaussian number density distribution of
unit norm and RMS radius $\langle\vec x^2\rangle=3r^2_0$,
\be
\rho_N(\vec x-\vec x_i)
 \ =\ \left({1\over2\pi r_0^2}\right)^{3/2} e^{-(\vec x-\vec x_i)^2/2r_0^2}.
\label{rhoN}
\ee

In the present version of our model, particle production or
annihilation is not yet included. In principle, it is no essential
problem to implement them, once an appropriate probability density is
at hand. This remains for future work.

\begin{table}
\caption{\em Quark and gluon color charges in the abelian approximation.
  The color names are
  chosen arbitrarily but consistently throughout this article. The color
  shortcuts are the ones used in Fig.~\ref{colors}.}
\label{ColorTable}
\begin{tabular}{cccc}
particle & color name & $q^3$ & $q^8$ \\
\tableline
$q$ & red $r$ & 1 & $1/\sqrt{3}$ \\
$q$ & green $g$ & -1 & $1/\sqrt{3}$ \\
$q$ & blue $b$ & 0 & $-2/\sqrt{3}$ \\
$\bar q$ & antired $\bar r$ & -1 & $-1/\sqrt{3}$ \\
$\bar q$ & antigreen $\bar g$ & 1 & $-1/\sqrt{3}$ \\
$\bar q$ & antiblue $\bar b$ & 0 & $2/\sqrt{3}$ \\
$g$ & red-antigreen $r\bar g$ & 2 & 0 \\
$g$ & red-antiblue $r\bar b$ & 1 & $\sqrt{3}$ \\
$g$ & green-antiblue $g\bar b$ & -1 & $\sqrt{3}$ \\
$g$ & green-antired $g\bar r$ & -2 & 0 \\
$g$ & blue-antired $b\bar r$ & -1 & $\sqrt{3}$ \\
$g$ & blue-antigreen $b\bar g$ & 1 & $\sqrt{3}$
\end{tabular}
\end{table}

%%%%%%%%%%%%%%%%%%%%%%%  WHITE CLUSTERS %%%%%%%%%%%%%%%%

\vvs
\subsection{Hadrons as Irreducible White Clusters}
\label{IrreducibleWhiteClusters}

\vs
If a quark-gluon plasma is formed in an ultrarelativistic
heavy-ion collision, the droplet as a whole has to be a
color-$SU(3)$ singlet.
Since our numerical model respects only an abelian
$U(1)\times U(1)$ subgroup of the $SU(3)$ symmetry group,
an ``$SU(3)$ color singlet'' is represented simply by a
``white'' cluster, i.~e. one that is color
neutral with respect to both the $\lambda^3-$
and the $\lambda^8-$charge (cf. Fig. \ref{colors}).
An $SU(3)$ color singlet is then built ``automatically''
by the presence of classical $A_{\mu}^3$ and  $A_{\mu}^8$ fields,
as solutions of (\ref{YangMills}).
Our simulation therefore starts off from a large but white
quark-gluon plasma droplet.

In nature, the hadronization process then divides the initial
large cluster into a number of hadrons that are themselves separate
color singlets. Correspondingly, our simulation is supposed to end up in
a final state of small, white clusters that can be interpreted as
hadrons. In this section, we investigate and classify the sorts of
small clusters that we expect to find in the final state.

Imagine a small, white cluster of color charges.
Suppose it cannot be subdivided into smaller but still white
subclusters. 
Then, flux tubes bind the cluster together, forcing it 
to rotate or oscillate in a ``yoyo'' mode \cite{LOH,PHDLOH}. Thus it
will emit $\sigma$ waves, which cool the cluster down. 
Finally, all the color charges are 
resting with respect to each other, being centered at the same 
point in space. If we choose the same color charge distribution 
width for all particle species, the final cluster's charge density 
will vanish identically; therefore, all fields will take on their
repective vacuum values in this cluster.

\begin{figure}
\centerline{\psfig{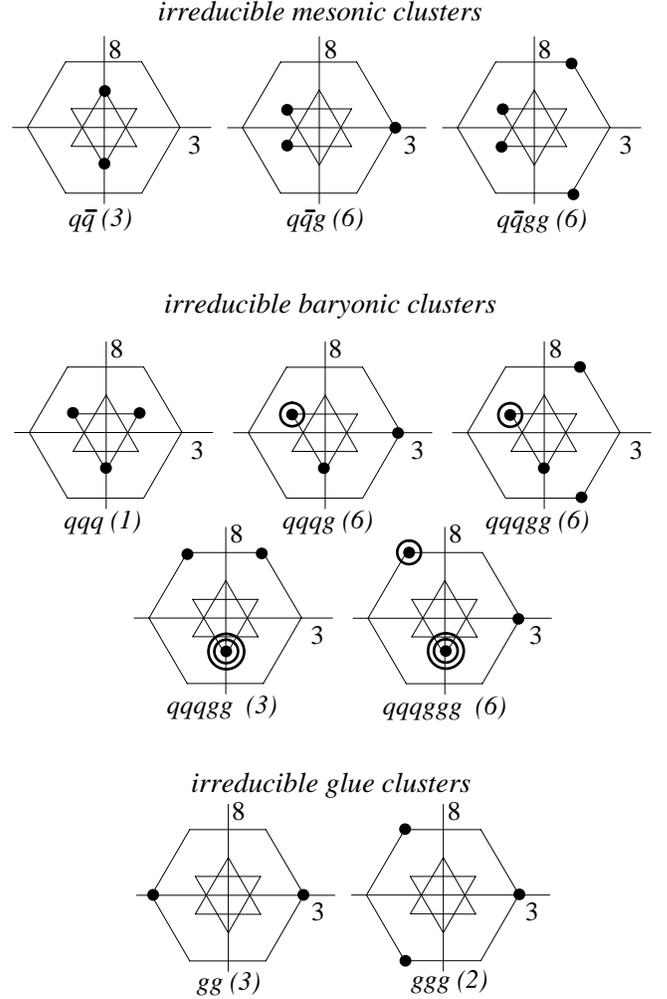}}
\caption{\em Classification of irreducible white clusters in the
color diagram of figure \ref{colors}. Points and circles indicate the
occupation number of each color state. Numbers in brackets indicate the
color degeneracy of the diagram for each quark flavor combination.}
\label{clusters}
\end{figure}

Because of the vanishing of all fields, two such clusters that 
happen to be close to each other can neither attract nor repel, 
and a small relative momentum suffices to separate them spatially. 
Consequently, we may
conclude that {\em the final state of our simulation must be a
collection of irreducible white clusters}, 
i.~e. white clusters that cannot be divided into
white subclusters. 
It turns out that only a finite number of possible
color combinations can form irreducible clusters, which enables
us to classify them, according to Fig.~\ref{clusters}, 
as mesonic, baryonic and pure glue ball states. 
The latter occur as mesonic resonance states in the experimental
observation.

We can turn this argument around, asking what clusters of quarks and
gluons in our model correspond to hadrons. A hadron is supposed to
be white but strongly bound. This leads us again to the notion of an
irreducible white cluster. If the hadron is in its ground state,
the colored constituents are all lying on top of each other, creating
a local color charge density that is vanishing everywhere.
Consequently, a cold hadron in our model
is completely field-free, and its mass
equals simply the sum of the masses of the constituents, without any
interaction contribution. E.~g., look at the $(qqq)$ baryon displayed
in Fig.~\ref{clusters}. The three quarks carry the colors red, green,
and blue, in the meaning of Fig.~\ref{colors}. This baryon is certainly
white, and it is also strongly bound: if one of the quarks, 
say the blue one ($q^3=0$, $q^8=-2/\sqrt{3}$), 
tried to escape, a flux tube of a strong $D^8-$field would build up,
pulling the fragments together with a force of 1 GeV/fm. In the ground
state, the three color charges sit right on top of each other, creating a
locally vanishing color charge density. In this case, all fields take on
their respective vacuum values, and the cold baryon possesses a mass $3 m_q$,
where $m_q$ is the quark mass.

When studying hadronization processes, our program
checks at each timestep whether there are irreducible white clusters
in the system. A ``cluster'' in the computational sense is a set of
color charges located in a common $\sigma$ field bag
(defined here as a connected spatial domain with
$\kappa>\frac{1}{2}$). The bag vanishes for cold clusters in our
model (see Sec.~\ref{IrreducibleWhiteClusters}), so if a
set of charges is outside any $\sigma$ field bag but very close in phase
space, we treat it as a cluster as well. If the cluster is irreducibly
white, i.~e. possesses one of the color structures displayed in
Fig.~\ref{clusters}, it is considered to be a new hadron.
The program then classifies the hadron according to its color
structure, computes particle and field
contributions to momentum and energy, removes the charges and field
imprints from the simulation as well as possible, and generates a
hadron, i.~e. a white, non-interacting object that just proceeds on a
straight path and may even leave the lattice.

A short side remark is in order:
In the investigation of the chromodielectric model, many authors (including the
original inventors of the model) added a term 
\be
{\cal L}_{p\sigma}= -\sum_{f}\bar q^f g_S\sigma q^f 
\label{gsCoupling}
\ee
in the Lagrangian, coupling the scalar quark density directly to
the $\sigma$ field. This generates a high constituent quark
mass outside a bag, which causes a high-density
phase transition to occur in the constant field approximation
(see, e.~g., \cite{BIRSE,MCGOVERN}). The coupling term also
establishes bound solitonic states \cite{LEE,VETTER,PHDVETTER,WILETSBOOK} with 
vanishing color charge density. Here we do not include 
the term (\ref{gsCoupling}) in the Lagrangian (\ref{Lagrangian}). 
If we did, the irreducible, final clusters of a
simulation run would not be field-free but rather generate a local
dip in the $\sigma$ field. This would generate a short-range 
attractive force between the white clusters: they would forever 
stick together, cool down by the emission of $\sigma$ waves, and 
end up as a single large and cold object, spoiling the 
hadronization process we want to model. 

In the model flavor presented here, there is neither an attractive nor
repulsive classical force between cold (field-free) hadrons. 
It is interesting to note in this context that Koepf et
al. \cite{KOEPF} have found 
a soft and short-ranged $NN-$repulsive potential in a quantum 
mechanical investigation within certain approximations of the very
same model.

%%%%%%%%%%%%%%%%%%%%%%%  SOLUTION OF THE FIELD EQUATIONS %%%%%%%%%%%%%%%%

\vvs
\subsection{Numerical Solution of the Field Equations}
\label{NumericalSolutionOfFieldEquations}

Let us recall the equations of motion obtained
from the Lagrangian (\ref{Lagrangian}) 
with the classical particle term (\ref{ClassicalParticleLagrangian}).
The quarks and charged gluons feel a Lorentz force coming from the 
color field,
\be
 m_i{d^2\vec x_i\over dt^2}
 \ =\ q_i^a\left(\vec E^a+{d\vec x_i\over dt}\times \vec B^a\right).
\label{LorentzForce}
\ee
The $\sigma$ field equation, rewritten from
(\ref{sigmaFieldEquation1}), reads
\be
{\partial^2\sigma\over\partial t^2}
 \ =\ \nabla^2\sigma-U'(\sigma)
 -{\kappa'(\sigma)\over 4}\left((\vec E^a)^2-(\vec B^a)^2\right).
\label{sigmaFieldEquation}
\ee

The $\sigma$ field is stored on a $125^3-$sized lattice with a volume
of $(25 {\rm fm})^3$, and both the particle and $\sigma$ field
equations of motion are solved by a simple explicit timestep method 
(staggered leapfrog, step-length $0.04$ fm). 

The color fields obey the classical Maxwell equations,
\ba
 \vec\nabla\cdot\left(\kappa(\sigma)\vec E^a\right)&=&\rho^a \label{GaussLaw} \\
 \vec\nabla\times\vec E^a&=&-{\partial\vec B^a\over\partial t}
  \label{InductionLaw} \\
 \vec\nabla\cdot\vec B&=&0 \label{CurlFreeLaw} \\
 \vec\nabla\times(\kappa(\sigma)\vec B)&=&\vec j^a
  +{\partial\kappa(\sigma)\vec E^a\over\partial t} \label{AmpereLaw}
\quad,
\ea
with charge and current density given by (\ref{ClassicalColorCurrent}).
These equations cannot be treated by an explicit timestep method.
The reason is that the Gauss law (\ref{GaussLaw}), being a constraint
equation, contains no time derivative and therefore constitutes an implicit
condition that has to be satisfied at each timestep separately. 

To simplify the problem, we neglect the $\vec B^{3,8}-$fields in our
simulation (and thereby neglect the corresponding color waves).
Although breaking Lorentz invariance, it is not as bad an
approximation as it may seem: it is exact for static charge configurations 
like in the computation of the string tension (Fig.~\ref{stringtension});
it is also exact for yoyo-like excitation modes of flux tubes in
their rest frame\cite{PUFF}. 
Furthermore, the color confinement mechanism explained in Sec.~
\ref{ConfinementMechanism} is not broken by this approximation. Finally, 
the color field waves cannot mediate forces or transfer energy between
hadronic bags because the Poynting vector $\kappa\vec E\times \vec B$
vanishes anyway in the physical vacuum. So the neglection of $\vec B$ 
amounts to a neglection of some inner degrees of freedom of clusters 
that are nonessential for the confinement transition.

Neglecting $\vec B^a$, it is clear from (\ref{InductionLaw})
that $\vec E^a$ can now be derived from a single potential,
\be
 \vec E^a\ =\ -\vec\nabla\phi^a.
\label{PotentialDefinition}
\ee
The Gauss law (\ref{GaussLaw}), rewritten in terms of $\phi^a$, reads
\be
 \nabla\cdot\left(\kappa(\sigma)\nabla\phi^a\right)\ =\ -\rho^a.
\label{PoissonEquation}
\ee
This is still a constraint equation, modelling the constraint of quark
confinement. It constitutes a boundary value problem that has to be 
solved at each timestep of the simulation. This has to be done 
carefully since the color electric flux is an essential ingredient of the 
confinement mechanism. 

Note that (\ref{PoissonEquation}) is singular in regions where
$\kappa$ vanishes; this is a major obstacle to numerical treatment.
We overcome it in practice by choosing a nonzero but small value 
for $\kappa_{vac}$ (see Sec.~\ref{ModelParameters}). 
Still, (\ref{PoissonEquation}) is nearly singular
and correspondingly difficult to solve.
A well-tuned successive overrelaxation scheme (SOR) on a fixed grid
with $N^3$ points, possessing at the very best a time complexity of
${\cal O}(N^4)$ \cite{NUMREC}, does not suffice here. 
Fourier transformation methods are excluded because of the
nonconstant coefficient function $\kappa(\sigma)$. So we choose a
fast solution algorithm based on the hierarchical basis multigrid 
method\cite{BANK}. We are using
a tetrahedral mesh of finite elements and take piecewise quadratic
basis functions for the Ritz-Galerkin variational ansatz.
During the multigrid iterations, the mesh is adaptively refined by 
bisection\cite{TRAXLER97}, which leads to block-diagonal matrices on
each level of the multigrid scheme\cite{MITCHELL}. Refinement is 
performed only in those regions where the solution function 
varies with short wavelengths, e.~g. close to the charges. 
This guarantees a certain precision of the solution $\phi^a$ 
independently of the structure of the sources $\rho^a$ or the 
coefficient function $\kappa(\sigma)$. 

There is a problem related to image charges. Imposing Dirichletian
boundary conditions on the boundary surface of our cubic domain, any
color charge that comes close to the boundary ``sees'' an image charge
reflected at the boundary surface. In fact, even flux tubes between a
charge and its image charge can be created, which yields a particularly
spectacular picture if a single color charge is put in the model world.
Image charges, however, are unphysical because the domain boundary
is unphysical --- real-world space is infinite.
Fortunately, there is a simple trick to remove image charges: we make
the cubic domain of the $\phi^a$ field much larger --- $(40 {\rm fm})^3$ ---
than the $(25 {\rm fm})^3$ lattice of the $\sigma-$field. Since no color
charge can leave the inner lattice (it would go into a
region where $\sigma=\sigma_{vac}$), it will never get close enough
to the domain boundary to feel forces from its image charge.
The adaptive mesh refinement method allows us to enlarge the
computational domain like this (a factor 5 in its volume!) 
without an even noticeable increase of computing time.

Typical meshes resulting from the refinement process contain 
about $10^4$--$10^5$ simplices (tetrahedra) on about 20 levels. The
Ritz-Galerkin ansatz decays on the various levels into about 
$10^4$ small linear systems of equations, each linear system involving 
about 40 nodal field values. 
This huge data structure has to be rebuilt from scratch in each
timestep, since $\kappa(\sigma)$ is changing. To save computing time, 
caching of matrices and lazy evaluation techniques prove to be
extremely useful.

The whole process of solving the Gauss law takes about 30 steps, in
which adaptive refinement alternates with relaxation (solution) of the
underlying equation. In each such relaxation run, the hierarchical
multigrid method relaxes the Ritz-Galerkin equations in some
$10$--$30$ V-cycle iterations; their convergence is further
accelerated by an Aitken transformation method applied to all nodes.

The CPU time thus needed for
one solution of (\ref{PoissonEquation}) varies between $15s$ and
$200s$ on a Pentium Pro 200MHz machine, depending mainly on the
structure of the source function $\rho^a$ and $\kappa$, which in turn
depend mainly on the number of charges present.

The quality of the solution is such that we never observe free color
charges in actual simulation runs, and even the interaction between
larger clusters that carry nonzero total color is strong enough
that they do not separate.

%%%%%%%%%%%%%%%%%%%%%%% STATICAL PROPERTIES OF FLUX TUBES %%%%%%%%%%%%%%%%

\vvs
\section{Statical Model Properties}
\label{StaticalModelProperties}

Before studying dynamical properties, like the time evolution of a
quark-gluon plasma, in the present model, let us briefly discuss
our choice of model parameters, which is based solely on certain
desired time-independent observables, namely the hadronic mass
spectrum, and color flux tube properties.

%%%%%%%%%%%%%%%%%%%%%%% CHOICE OF MODEL PARAMETERS %%%%%%%%%%%%%%%%

\vs
\subsection{Choice of Model Parameters}
\label{ModelParameters}

\vs
We choose the same charge distribution radius \mbox{$\sqrt{\langle\vec
x^2\rangle}=0.7$ fm} for all the various
particle flavors. The model possesses in this case a particularly
simple ``hadron spectrum''.
Cold (unexcited) hadrons are white (color neutral) combinations of several
particles centered at the same location. Their color charge distribution
vanishes everywhere, therefore the fields all take on their vacuum values. Thus,
the mass of such a hadron is simply the sum of the masses of its
constituents. 
In order to obtain roughly the proper hadronic mass scales, we have
chosen a nonstrange quark mass of 400 MeV, yielding a mass of 800 MeV
for the $\rho-$meson and 1200 MeV for the nucleon. The other masses
are all given in table \ref{FittedMassTable}. This table exhibits a
reasonable agreement between model and experimental numbers;
fine-tuning could certainly improve this even further.

\begin{table}
\caption{\em Quark and gluon effective masses,
  and corresponding hadron masses
  [Particle Data Table]. All masses are measured in MeV.
  The symbol $q$ denotes an up or down quark,
  $s$ a strange quark, and $g$ a gluon.}
\label{FittedMassTable}
\begin{tabular}{cccc}
particle & model mass & exp. mass & error \\
\tableline
$q$ & 400 & - & - \\
$s$ & 550 & - & - \\
$g$ & 700 & - & - \\
${1\over2}(N+\Delta)$\ $(qqq)$  & 1200 & 1086 & +10.5\% \\
% $\Delta$\ $(qqq)$ & 1200 & 1232 & -2.5\% \\
$Y$\ $(qqs)$ & 1350 & 1385 & -2.5\% \\
$\Xi$\ $(qss)$ & 1500 & 1530 & -2.0\% \\
$\Omega$\ $(sss)$ & 1650 & 1672 & -1.3\% \\
$\rho$\ $(q\bar q)$ & 800 & 770 & +3.9\% \\
$\omega$\ $(q\bar q)$ & 800 & 782 & +2.3\% \\
$K^*$\ $(q\bar s)$ & 950 & 892 & +6.5\% \\
$\Phi$\ $(s\bar s)$ & 1100 & 1020 & +7.8\% \\
$gg$ & 1400 & 1400\ ? 1& - \\
\end{tabular}
\end{table}

%Such a constituent quark model without any interaction
%contribution to hadronic ground state energies can never
%successfully describe the low masses of the light, strongly bound hadrons,
%like $\pi$ or $K$ mesons, or nucleons.
%It is interesting in this context that the experimental masses
%of the first excitations can be tuned in very well with roughly chosen
%quark masses (see table \ref{FittedMassTable}).
%Of course, this fit is not meant as a serious model of hadrons.
%Since we neglect spin degrees of freedom, one could also take the
%viewpoint that the $(qqq)$ hadron of our model should correspond to
%a mass roughly equal to the average mass of the nucleon and $\Delta$.
%This is a little below our model value.
%We feel, however, that our simulation of classical color dynamics
%is not too sensitive to the precise choice of particle masses.

The parameters of the quartic potential $U(\sigma)$ and the dielectric
function $\kappa(\sigma)$ are harder to fix since there is no well-known
physical particle associated with $\sigma$. However, if we interpret $\sigma-$waves as
pure glueballs with a mass of about $1400 MeV$, we obtain
\mbox{$U''(\sigma_{vac})=(1400 MeV)^2$}. Furthermore, the bag constant
\mbox{$B=(150 MeV)^4$}, and \mbox{$\sigma_{vac}=0.31 {\rm fm}^{-1}$}
are chosen in order to get a low bag surface tension --- which is
easier to handle numerically --- while still having a region with
$U''(\sigma)<0$, which is a necessary condition for a
first order color deconfinement phase transition to occur.
Our particular choices are motivated by studies of solitonic states
in the model \cite{VETTER,PHDVETTER}. Finally, we use
\be
U(\sigma)\ =\ B + a\sigma^2 + b\sigma^3 + c\sigma^4
\label{Usigma}
\ee
with $a=6.184 {\rm fm}^{-2}$, $b=-80.72 {\rm fm}^{-1}$, $c=163.1$,
$B=(150 \mbox{MeV})^4$. For $\kappa$, it turns out that the simulation results
are not very sensitive to its precise shape. As described in the
last section, its only purpose is to couple the vacuum pressure generated by
$U(\sigma)$ to the color flux tube. For doing this, the $\kappa$
function has to take on a value much smaller than one near
$\sigma\approx\sigma_{vac}$. Secondly, $\kappa$ has to rise quickly to
one where $U(\sigma)\approx B$, supporting our interpretation of $B$ as bag
constant. We use
\be
\kappa(\sigma)\ =\ {1\over\exp\left(\alpha
  \left({\sigma\over\sigma_{vac}}-\beta\right)\right)+1}
\label{kappasigma}
\ee
with $\alpha=5$ and $\beta=0.4$.

%Both functions $U$ and $\kappa$ are plotted
%in Fig.~\ref{Ukappa_numerical}.
%
%\begin{figure}
%\centerline{\psfig{figure=Ukappa_numerical.eps,width=90mm}}
%\caption{\em The functions $U(\sigma)$ and $\kappa(\sigma)$ used in our
%computer simulation.}
%\label{Ukappa_numerical}
%\end{figure}

\begin{figure}
\centerline{\psfig{figure=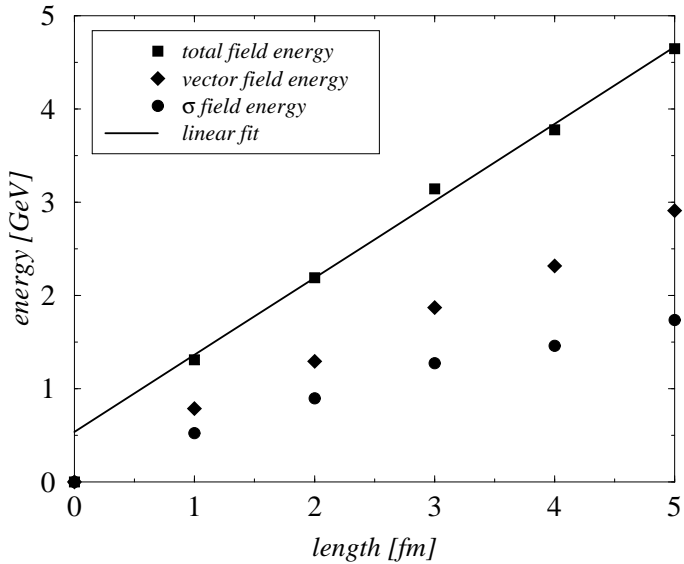,width=90mm}}
\caption{\em The $q\bar q$-flux tube energy ($\sigma$ and color field
contributions) plotted against the separation distance of the quarks.
The slope is about $1$ GeV/fm.}
\label{stringtension}
\end{figure}

Finally, the strong coupling constant \mbox{$\alpha_S=g_v^2/4\pi$}
is chosen such that the $q\bar q$
string tension equals about $1$ GeV/fm, a value that can be deduced from quarkonium
spectroscopy \cite{PERKINS}. Anticipating the results of the next section, we
show in Fig.~\ref{stringtension} that a value of $\alpha_S=2$ results in a
satisfactory numerical string tension.

%%%%%%%%%%%%%%%%%%%%%%% STATICAL PROPERTIES OF FLUX TUBES %%%%%%%%%%%%%%%%

\vvs
\subsection{Statical Properties of Flux Tubes}
\label{StaticalPropertiesOfFluxTubes}

\begin{figure}
\centerline{\psfig{figure=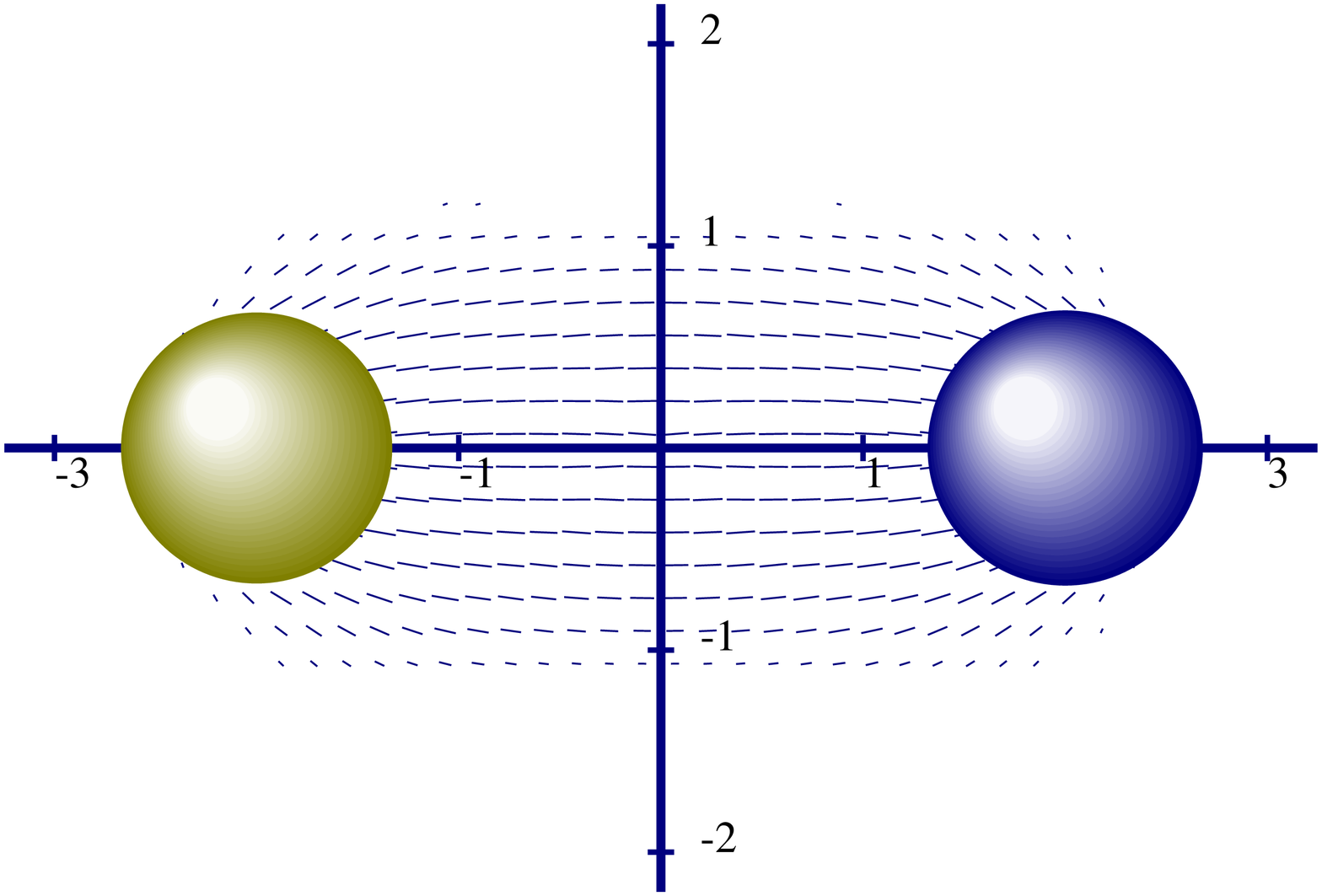,width=70mm}}
\centerline{\psfig{figure=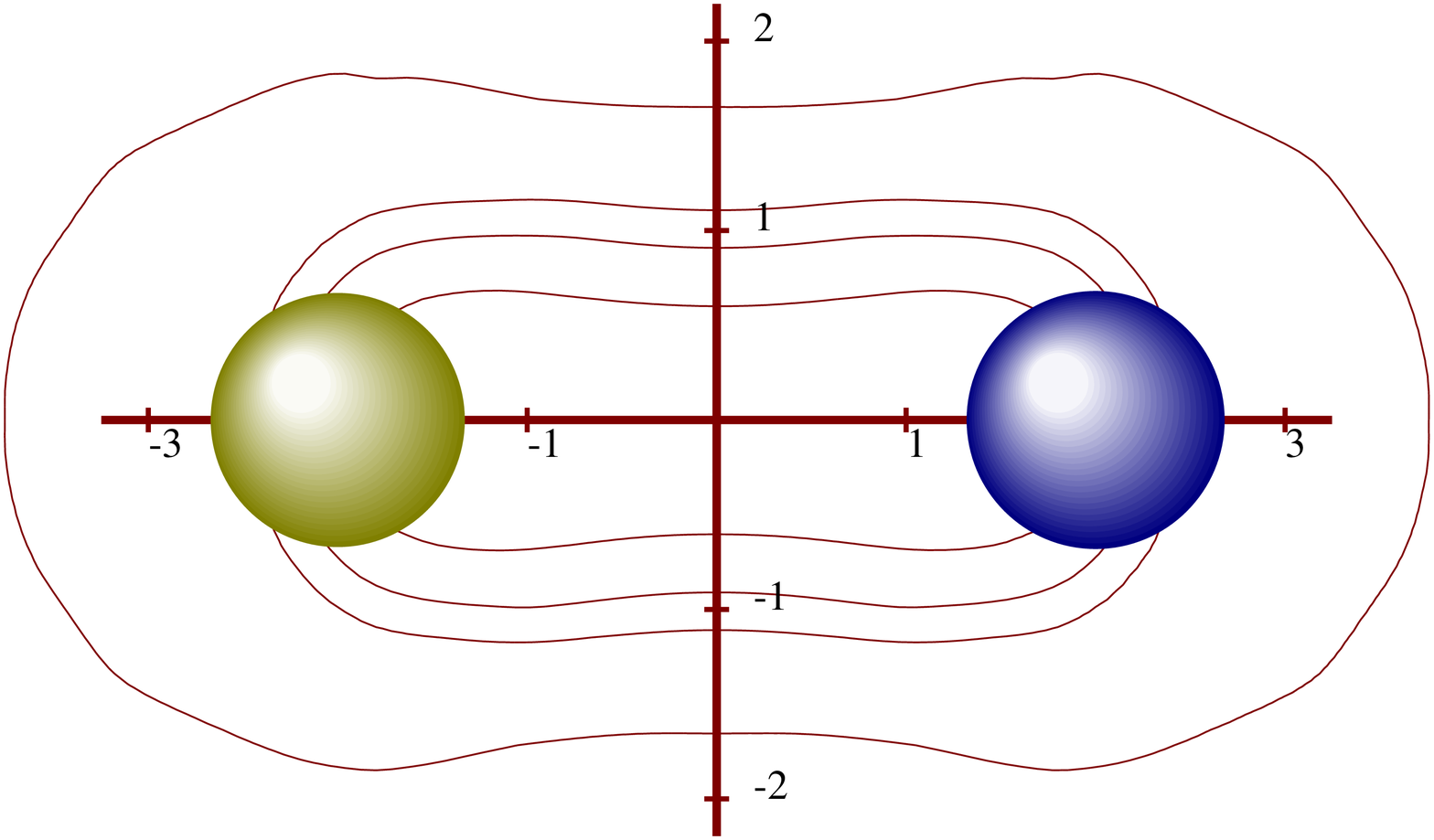,width=70mm}}
\centerline{\psfig{figure=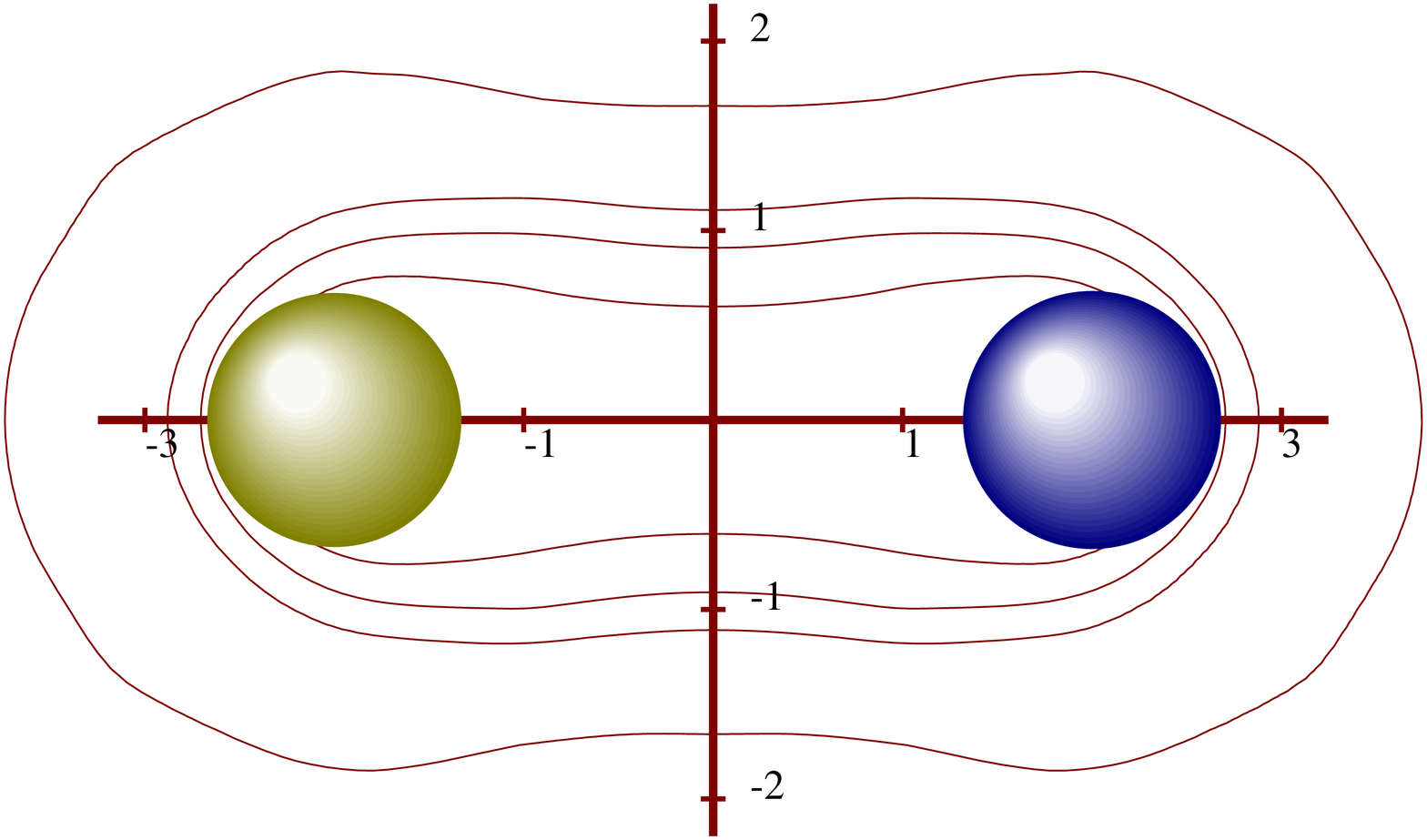,width=70mm}}
\centerline{\psfig{figure=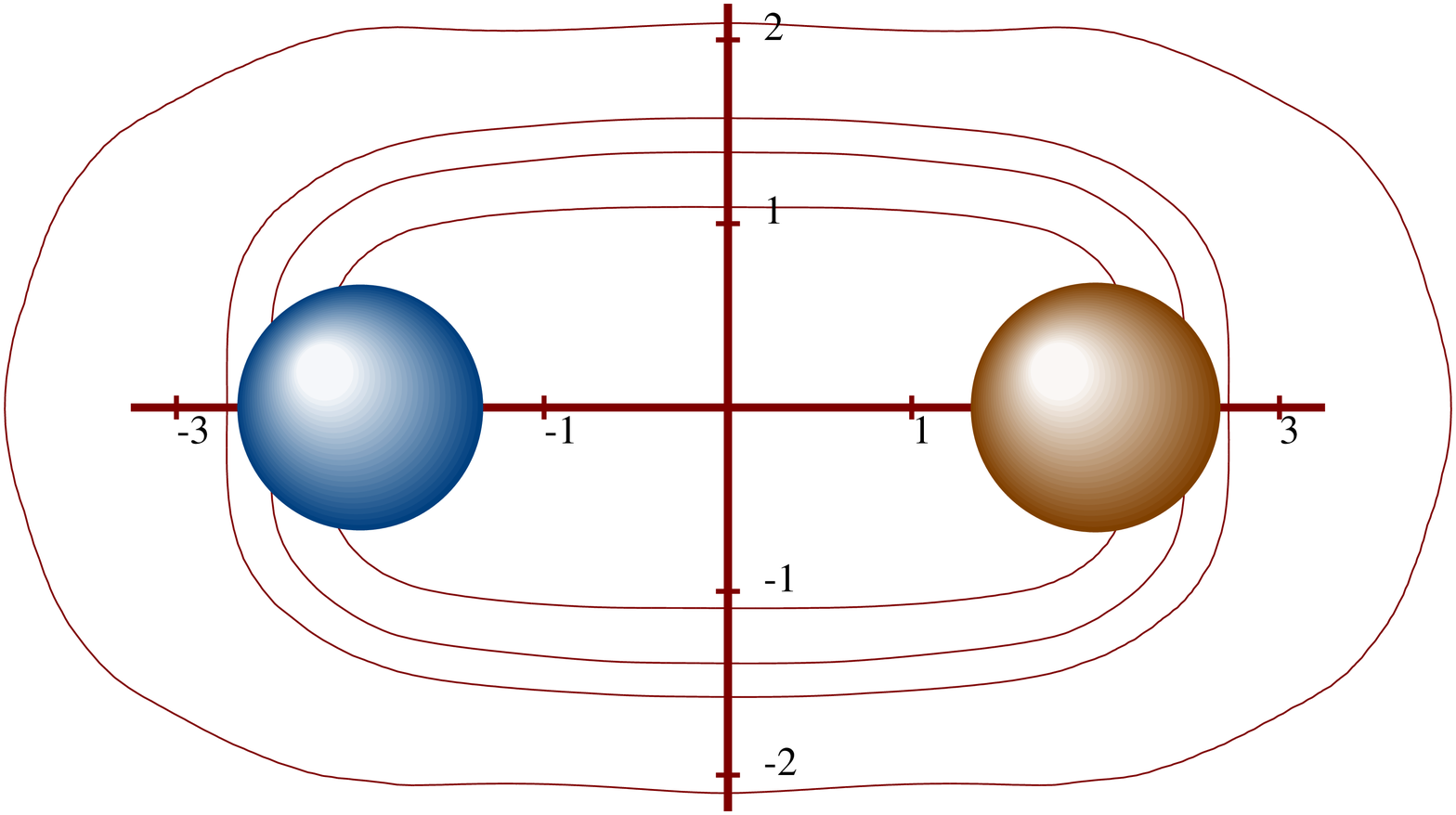,width=70mm}}
\caption{\em $(a)$: Vector field of a quark string of length
4 fm, $(b)$ its $\sigma$ field, $(c)$ modification caused by an
additional scalar coupling term (\ref{gsCoupling}) with $g_S=8$,
$(d)$ $\sigma$ field of a gluonic string. The contour lines
are defined by $\sigma=0$, $0.1$, $0.2$, and $0.3$
$(\sigma_{vac}\approx 0.31)$. Axis ticks are in units of 1 fm.}
\label{stringtypes}
\end{figure}

Of course, the string tension, plotted in Fig.~\ref{stringtension} 
and already discussed in the last section, is the flux tube property
most relevant to the hadronization process.
Also, all our model parameters are already fixed by the
considerations done in the last section. Nevertheless, we would like
to show pictures of color flux tubes and their fields.
We do this in Fig.~\ref{stringtypes}: Picture $(a)$ shows the color flux
$\vec D^8=-\kappa(\sigma)\nabla\phi^8$ between a blue quark and an
antiblue antiquark (in the nomenclature of Fig.~\ref{colors}).
The vector field is displayed in a plane containing
both quark centers --- the computation has of course been performed in
three dimensions. The spheres at both ends of the flux tube
represent the quarks; their radius indicates the RMS radius of the color
charge distributions ($0.7$ fm). It can be clearly seen that the color
flux is indeed nicely parallel, as would be expected from a string
configuration. The string radius turns out to be rather large ---
about 1 fm. In terms of our irreducible white cluster classification
of Fig.~\ref{clusters}, this string is an excited meson of the $q\bar q-$type.

Picture $(b)$ of Fig.~\ref{stringtypes} shows the $\sigma$
field for the same charge configuration.
The contour lines are drawn at $\sigma$
values of $0$, $0.1$, $0.2$, $0.3$ (from the inside outwards),
where $\sigma_{vac}\approx0.31$.
Remember that the only source term of the $\sigma$ field is given by
$\kappa'(\sigma)F^2/4$; thus, the vector field drawn in picture $(a)$
effects a field imprint that does not fully extend over the quarks.
Now we added a direct quark-$\sigma$ coupling term (\ref{gsCoupling}),
taking $g_S=8$ and approximating $\bar qq$ by the quark number density
given in (\ref{rhoN}). The result is picture $(c)$, differing from
$(b)$ only slightly, since the quark-$\sigma$ coupling
affects only the endpoints of the flux tube.

Lastly, picture $(d)$ shows the $\sigma$ field for a gluonic
string. The total color flux is larger here: since the gluons carry
each the sum of a quark and an antiquark color, the total gluon charge
$\sqrt{(q^3)^2+(q^8)^2}$
is larger by a factor of $\sqrt{3}$ in comparison with a quark charge
(see Sec.~\ref{MolecularDynamicsApproach}; for this picture, we took
the gluon colors $r\bar g$ and $g\bar r$). The total color flux
is therefore enhanced by the same factor (enforced by the Gauss law).
This results in a larger pressure of the flux lines,
which dig a significantly wider ``hole'' in the nonperturbative
vacuum. In terms of our irreducible cluster
classification scheme of Fig.~\ref{clusters},
this gluonic string is an excited glueball of type $gg$.

\begin{figure}
\psfig{figure=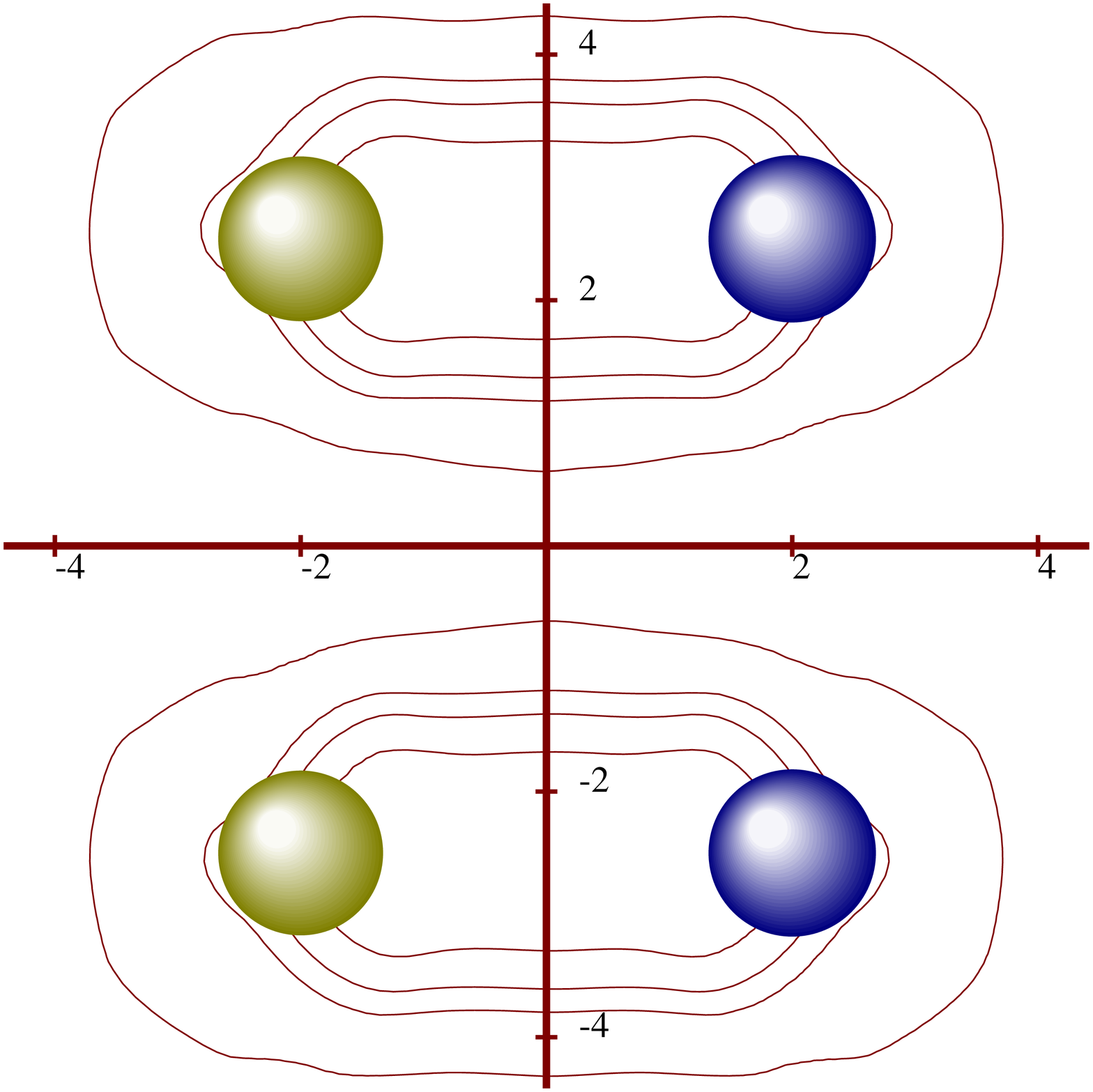,width=75mm}
\psfig{figure=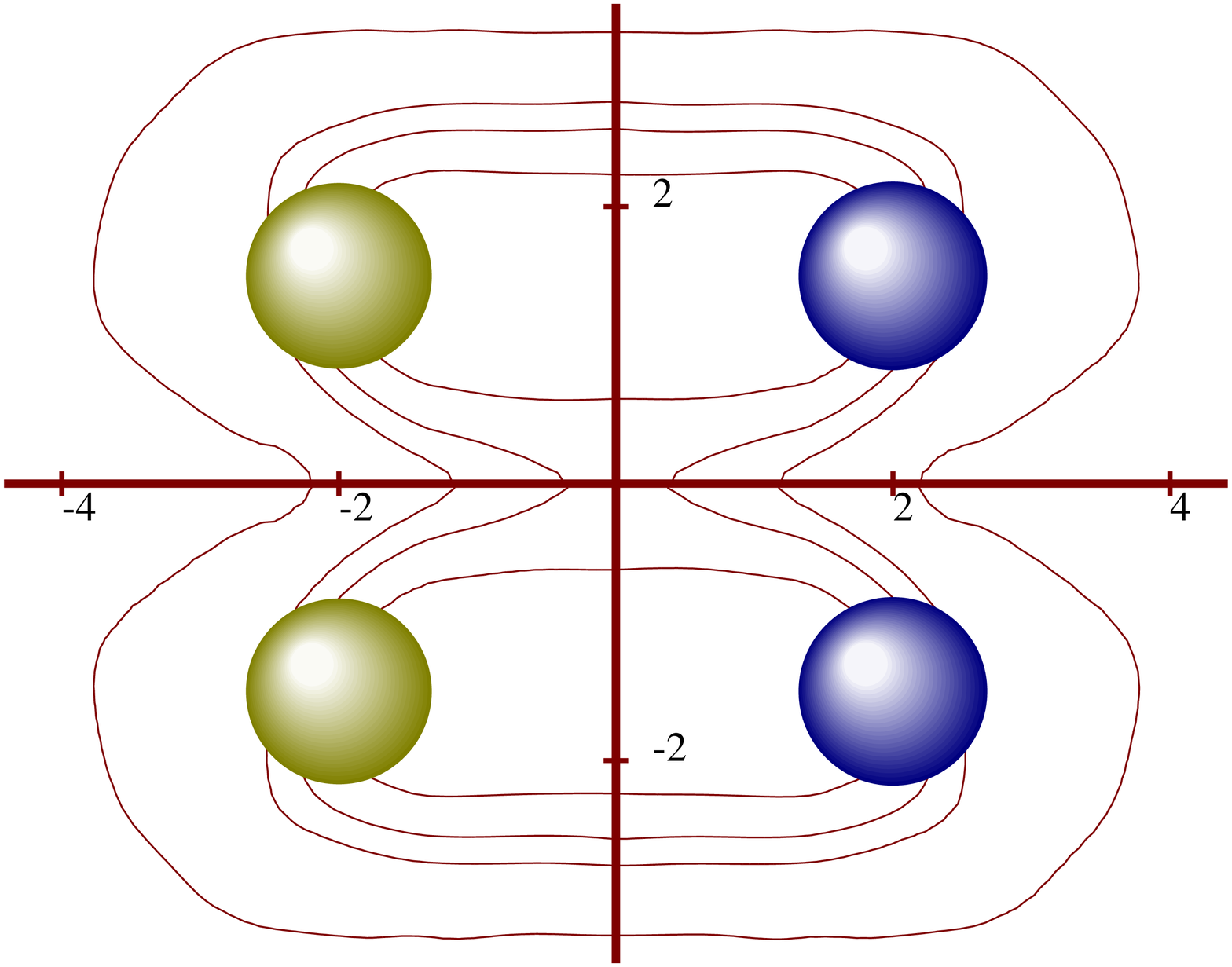,width=75mm}
\psfig{figure=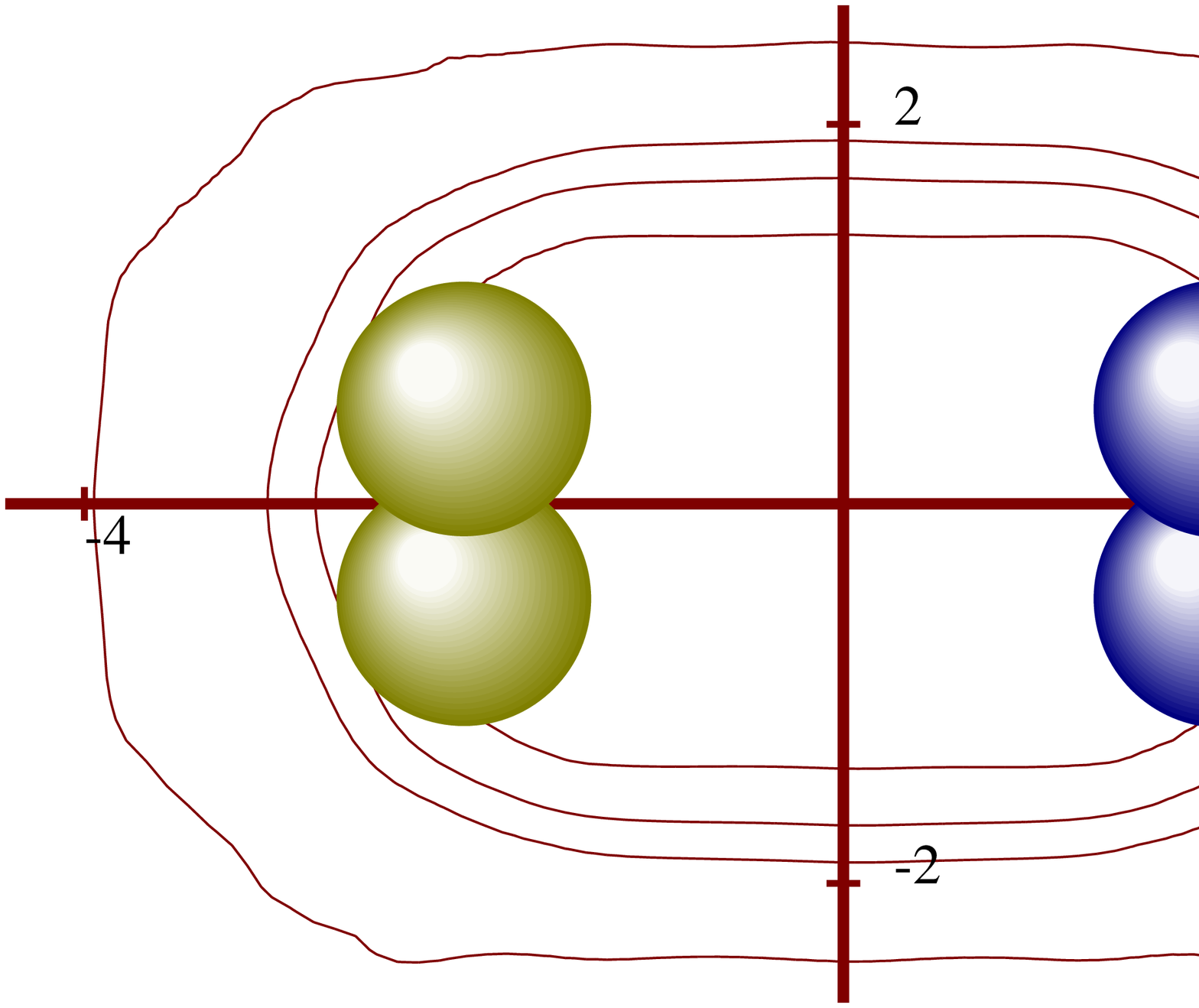,width=75mm}
\caption{\em Fusion of parallel flux tubes. The contour lines are at the
same $\sigma$ field values as in Fig.~\protect\ref{stringtypes}.}
\label{parstringfusion}
\end{figure}

\begin{figure}
\psfig{figure=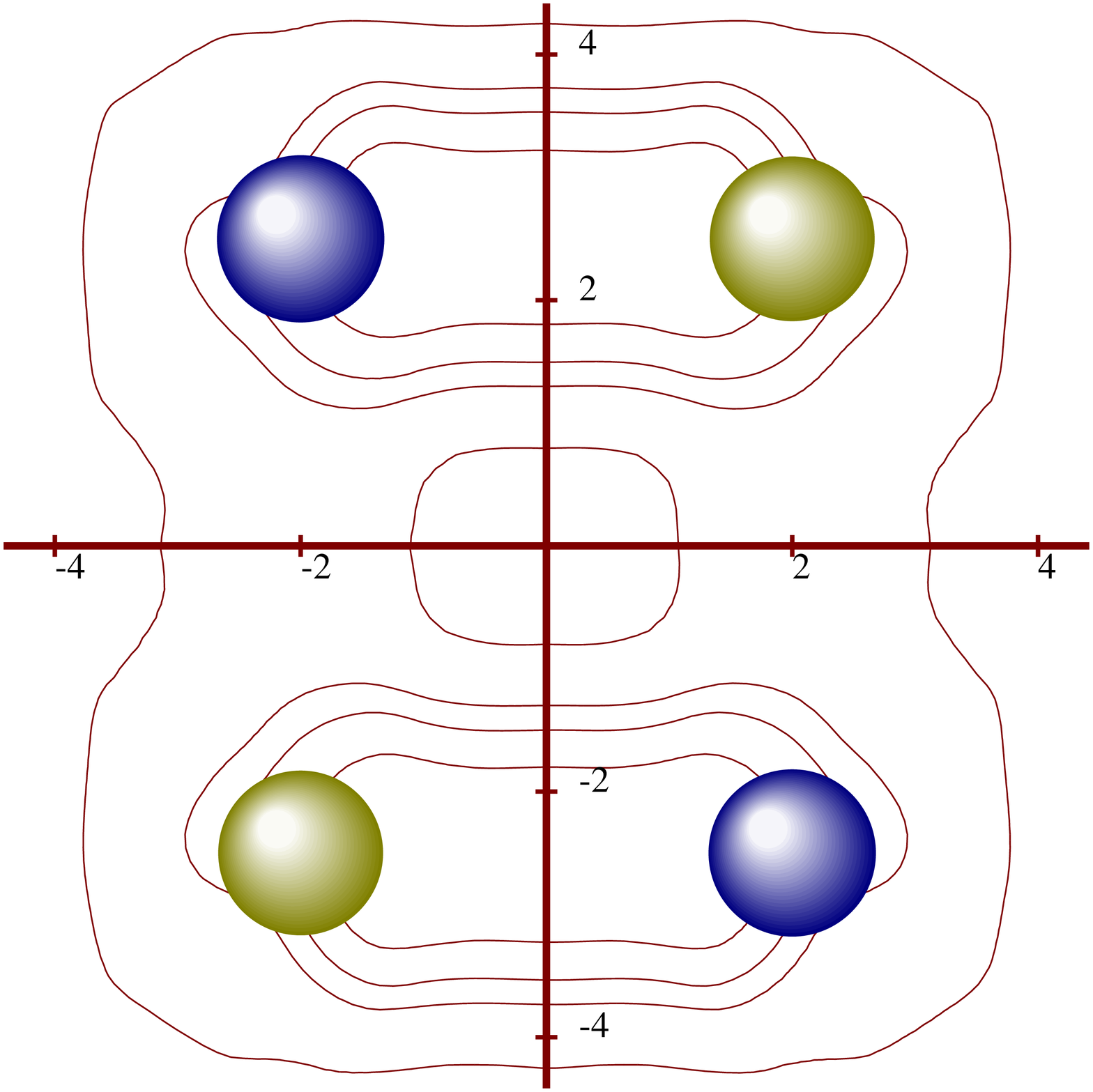,width=75mm}
\psfig{figure=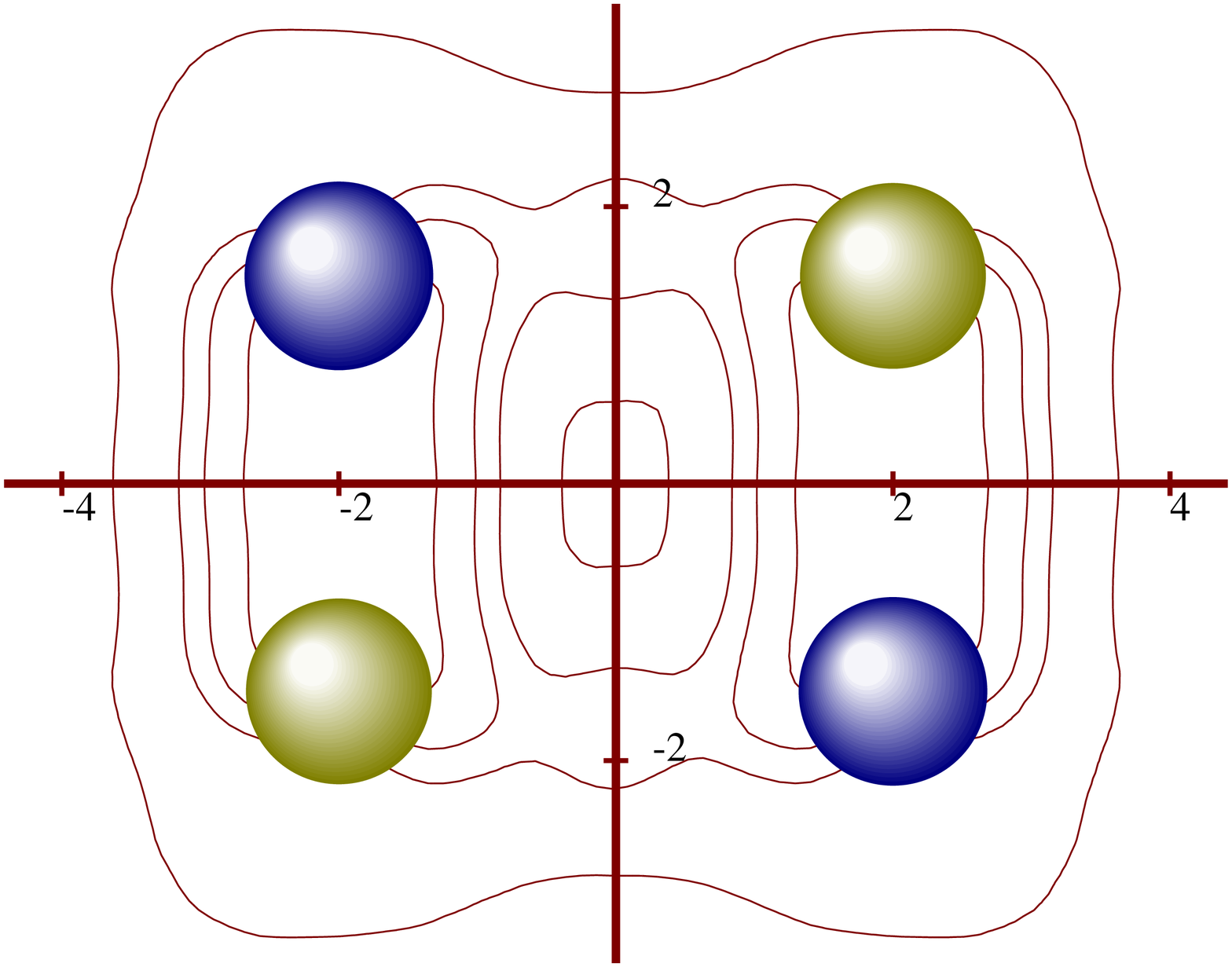,width=75mm}
\psfig{figure=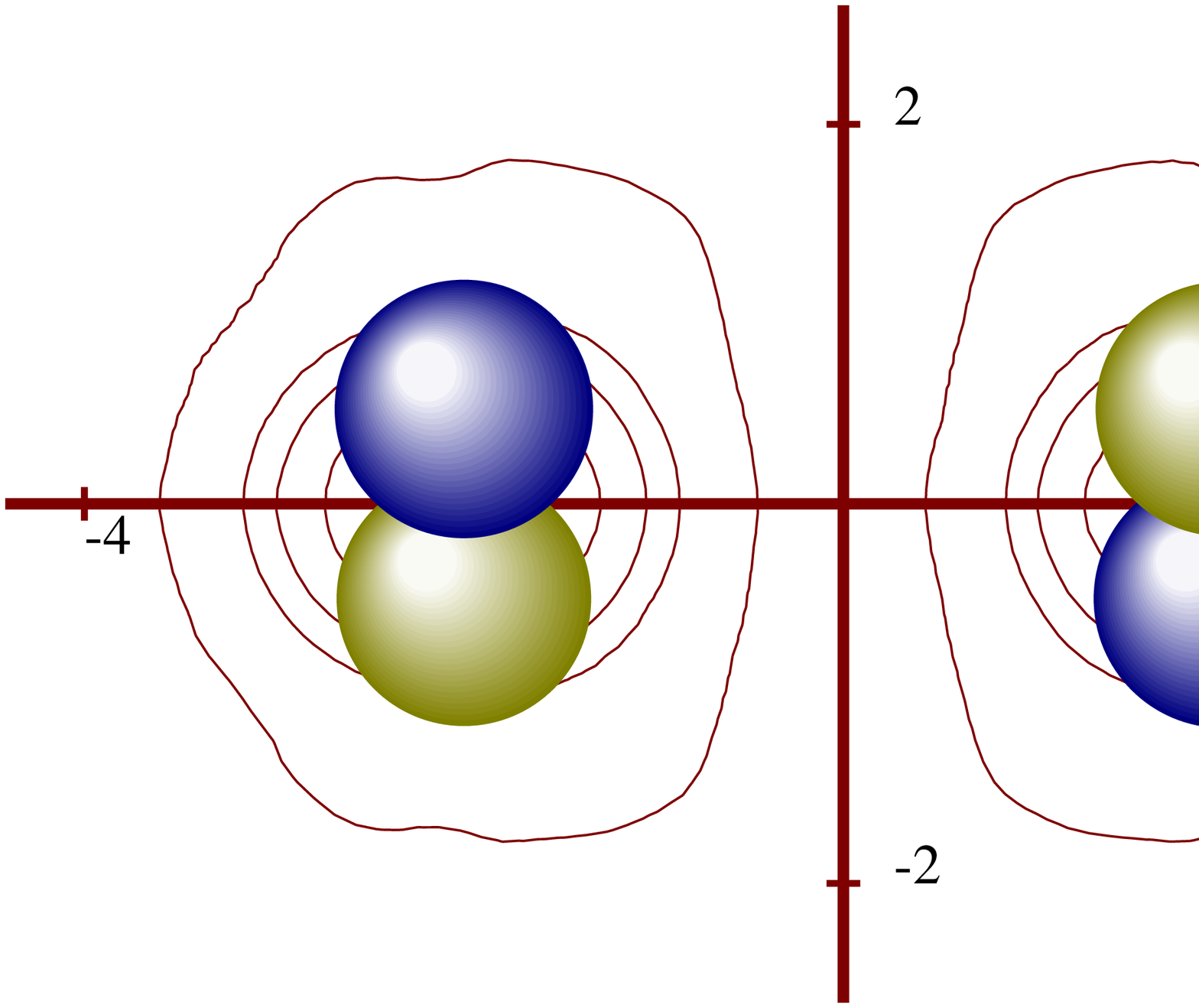,width=75mm}
\caption{\em Flux tubes flip in an antiparallel configuration.}
\label{antistringfusion}
\end{figure}

We can also use our program to investigate a possible string-string
interaction. The $\sigma$ field of parallel and antiparallel
configurations of two strings is plotted in
Figs.~\ref{parstringfusion} and \ref{antistringfusion}. 
For the parallel case (Fig. \ref{parstringfusion}), we see
no notable deformation of the strings as they approach each
other, while in the antiparallel case (Fig. \ref{antistringfusion}), 
the strings ``flip'', since the
field configuration of lowest energy (which is the static solution of the
field equations) is the one that connects only the closest quarks by 
flux tubes.

\begin{figure}
\psfig{figure=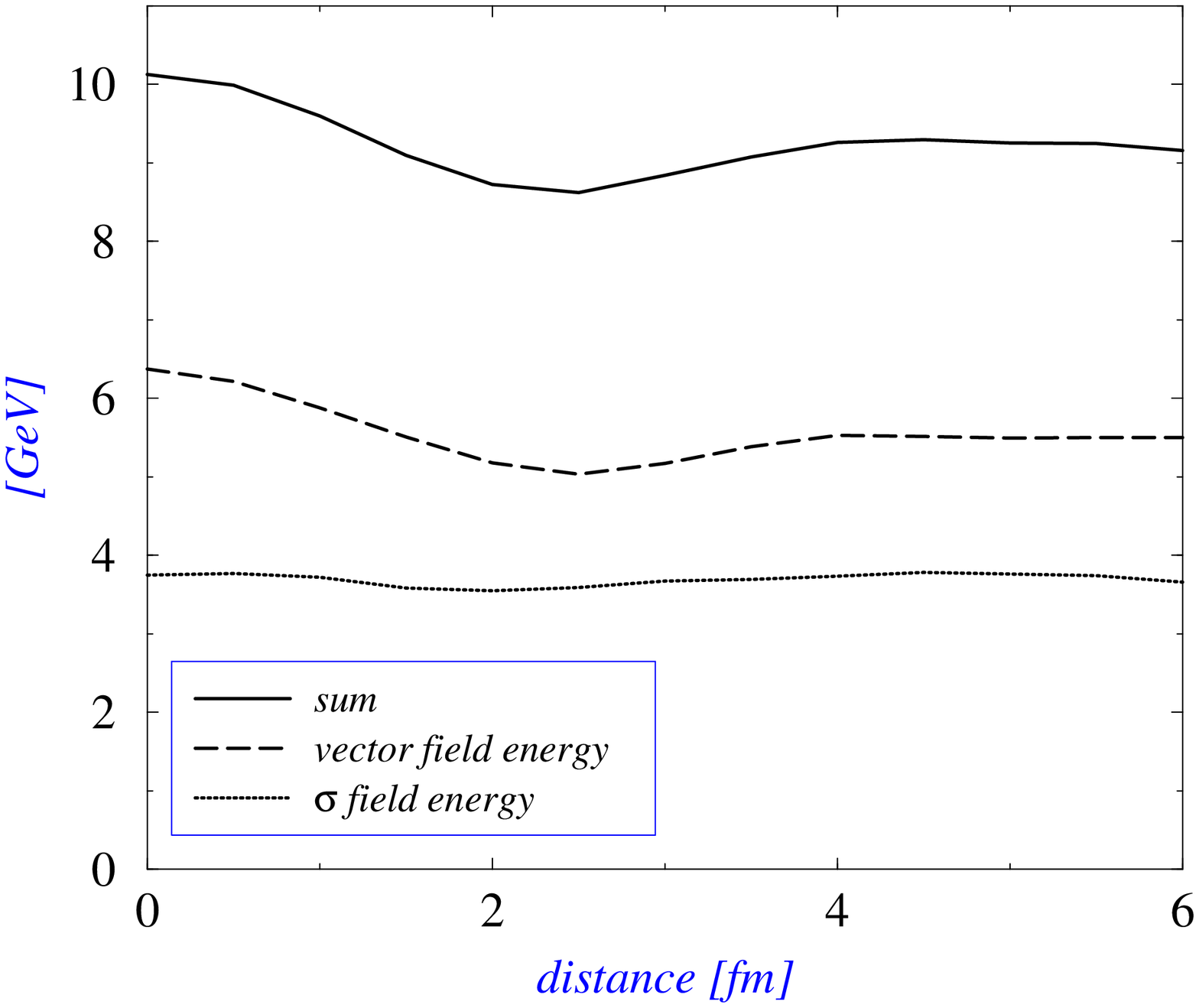,width=70mm}
\caption{\em String-string potential for the parallel case.}
\label{parstringpot}
\end{figure}

\begin{figure}
\psfig{figure=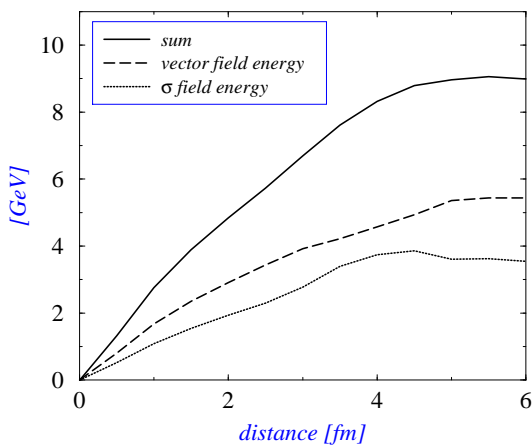,width=70mm}
\caption{\em String-string potential for the antiparallel case.}
\label{antistringpot}
\end{figure}

Figs.~\ref{parstringpot} and \ref{antistringpot} show the 
corresponding string-string potentials. 
For the parallel case (Fig. \ref{parstringpot}), the potential 
differs little from a constant, possessing a slightly attractive part 
(bag fusion) and a repulsive core (Coulomb repulsion). The potential of the
antiparallel configuration (Fig. \ref{antistringpot}) 
shows the expected stringlike (linear) behaviour as soon as the 
strings are flipped (below 4 fm).

%\begin{figure}
%\onecolumn
%\centerline{\psfig{figure=fusion_par_dist_5_sigma.eps,width=80mm}
%            \psfig{figure=fusion_anti_dist_5_sigma.eps,width=80mm}}
%\centerline{\psfig{figure=fusion_par_dist_3_sigma.eps,width=80mm}
%            \psfig{figure=fusion_anti_dist_3_sigma.eps,width=80mm}}
%\centerline{\psfig{figure=fusion_par_dist_1_sigma.eps,width=80mm}
%            \psfig{figure=fusion_anti_dist_1_sigma.eps,width=80mm}}
%\centerline{\psfig{figure=fusion_par.eps,width=70mm}
%            \psfig{figure=fusion_anti.eps,width=70mm}}
%\caption{\em Left: Fusion of parallel flux tubes. Right: Flux tubes
%flip in an antiparallel configuration. The contour lines are at the
%same $\sigma$ field values as in Fig.~\protect\ref{stringtypes}. The
%bottom picture in each column shows the corresponding string-string 
%potential.}
%\label{stringfusion}
%\twocolumn
%\end{figure}

%%%%%%%%%%%%%%%%%%%%%%% HADRONIZATION %%%%%%%%%%%%%%%%

\vvs
\section{Hadronization of a Quark-Gluon Plasma}
\label{HadronizationOfQGP}

The described simulation tool enables us to
follow the microscopic dynamics of hadronization.
However, in order to gain a qualitative insight, we 
restrict the first analysis to a few observables.
In the next section, we present our choice of hadronization
scenarios.
Then, we introduce some important observables
and explain our means of obtaining them from simulation data.
Finally, we discuss our results on mass distributions,
the evolution of the quark-gluon-hadron matter
composition, and analyze spectral properties.

%%%%%%%%%%%%%%%%%%%%%%%  INITIAL STATE %%%%%%%%%%%%%%%%%%%%%%%%%%%

\vvs
\subsection{The Initial State of the Simulation}

\vs
When carrying out a molecular dynamic simulation on the quark-gluon
level, one has to make compromises. Even using modern computers
and state-of-the-art numerical methods, we cannot simulate systems
larger than a few hundred of particles in a volume of
$(25 {\rm fm})^3$. The entire simulation run takes finally about a 200 fm,
corresponding to weeks of CPU time on a Pentium Pro 200 MHz system.

Due to the above mentioned restrictions we decided to simulate
only a central part of the heavy-ion collision region, including
about 400 quarks and gluons in an initial volume of about
100 fm${}^3$. In real $Pb+Pb$ collisons at CERN SPS, about 
$3000-4000$ particles are present in the quark matter phase.
We compare two different scenarios in order to gain a qualitative
insight about the hadronization: $(i)$ a fully stopped
quark matter at initial temperature $T_0=160$ MeV, and
$(ii)$ a hot quark-gluon plasma initialized on 
top of a Bjorken-flow stretching from coordinate-rapidity
$\eta_0=-0.5$ to $\eta_0=+0.5$.

The reaction volume is calculated assuming a radius
of the initial sphere of $R_0= 4$ fm or of the
initial cylinder of $R_0 = 4$ fm, respectively.
The initial energy density of $\varepsilon_0 = 2.5$ GeV/fm$^3$
is obtained from the total bombarding energy characteristic
for SPS. The initial numbers of quarks and gluons are chosen
according to non-relativistic Maxwell-Boltzmann statistics, 
using the respective masses and the initial temperature.
The ratios of various flavor quarks and gluons are determined
from equilibrium principles (law of mass action), being
proportional to
\be
N_i(0) \propto e^{-m_i/T_0} d_i \left( \frac{m_iT_0}{2\pi} \right)^{3/2},
\ee
with $d_i$ the spin, color and flavor degeneracy.
The resulting parton numbers are 133 light quarks, 40 strange quarks,
the same numbers of respective antiquarks, and 44 gluons for the
Bjorken event, and 196 light quarks, 52 strange quarks, and 57 gluons
for the full stopping event. The random Gaussian 
momenta (on top of the initialized flow pattern) satisfy
\be
\left\langle {p_{i,\mu}^2\over m_i} \right\rangle = T_0
\ee
for each particle species and momentum components $\mu = x, y$ or $z$.
We do not consider a nonzero quark chemical potential in the present
work; furthermore there is no net color and no net strangeness 
in the system.

A short side remark: A possible third scenario, a fully 
``transparent'' picture where the
particles are initially separated in phase space, leads to a rapid 
separation into unconnected parts (the spectators), leaving only
a small fraction of mini-clusters in the mid region.
Unfortunately, we cannot simulate too large
relative momenta at present, due to the computational restriction
to intermediate size simulation volumes. Some particles bounce
back from the lattice boundary and interfere artificially
with others still in the expanding fireball. This is unphysical, 
and we have to abandon our studies of this scenario at this point.

\vvs
\subsection{Identification of Observables}

\vs
The basis of our identification of observables is the
white cluster classification and hadronization mechanism
described in Sec.~\ref{IrreducibleWhiteClusters}.
The total momentum and total energy of white clusters can be
calculated from the individual four-momenta of its constituents
and from the field energy carried by the classical vector fields.
The mass of a cluster is defined by
\be
M^2_{{\rm clus}} = \left( E_{{\rm field}} + \sum_i E_i \right)^2 
-  \left( \vec{p}_{{\rm field}} + \sum_i \vec{p}_i \right)^2.
\ee
Here,
\ba
 E_{{\rm field}}&=&\int \! d^3x \, \left(
\frac{1}{2} \dot{\sigma}^2 + \frac{1}{2} (\vec{\nabla}\sigma)^2 
+ U(\sigma) + \frac{1}{2} \kappa(\sigma) \vec{E}^a \vec{E}^a \right)
\nonumber \\
 \vec p_{{\rm field}}&=&-\int \! d^3x \, \dot\sigma\vec\nabla\sigma.
\nonumber \\
 & &\qquad 
\ea

The mass distribution, $dN/dM$, tells the number of clusters
with a mass between $M$ and $M+dM$. The resolution $dM$ of
the mass binning is chosen to be $dM = 100$ MeV presently.
If $dM$ is too large, one cannot distinguish different
particles; if it is too small, the statistics becomes too low.
The above value is a compromise found by trial and error.

Further observables are the rapidity distribution $dN/dy$ with
\be
y = \frac{1}{2} \ln \frac{E+p_z}{E-p_z},
\ee
and the transverse momentum distribution $dN/dp_T$ with
\be
p_T = \sqrt{p_x^2+p_y^2}.
\ee
Collective flow and local temperature both affect these spectra:
a strong longitudinal flow causes a plateau in the $dN/dy$
distribution, while the slope of the transverse momentum
spectra --- if nearly exponential --- includes both thermal and
radially collective motion.

Finally, some global variables are of interest in order to
obtain an approximate equation of state from the molecular
dynamic simulation. A temperature-like quantity can be obtained
by fitting the non-relativistic Maxwell-Boltzmann distribution
to the momenta by inspecting
\be
T_{i,\mu} = \langle p_{i,\mu}^2 \rangle / m_i,
\ee
for a cluster of type $i$ with mass $m_i$, and momentum direction
$\mu=x,y$ or $z$. The approximate equality of these
temperature-like parameters for different spatial directions $\mu$
reflects isotropy of the pressure, while its equality between
different sorts of clusters $i$ chemical and thermal equilbrium.

It remains for future work to analyze the cluster
size distribution with its higher moments after collecting improved
statistics. This can shed some light onto the order of the
hadronization transition from the viewpoint of percolation
theory, as recently pointed out by H.~Satz \cite{SATZ-NEW}.

\begin{figure}
\onecolumn
\centerline{\psfig{figure=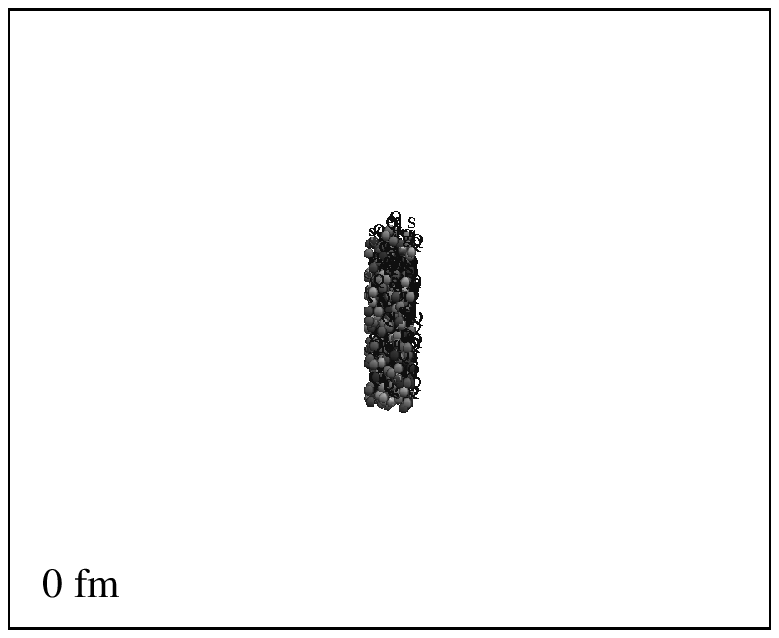,width=80mm}
            \psfig{figure=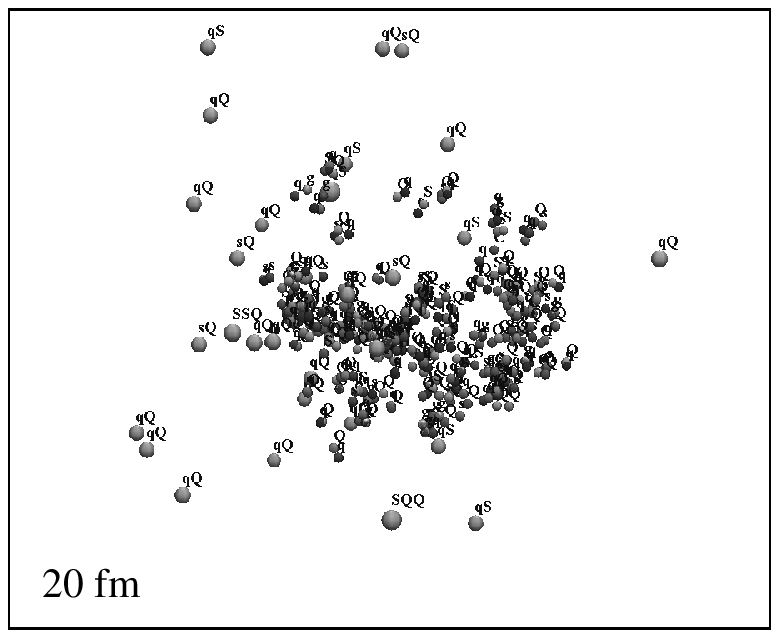,width=80mm}}
\centerline{\psfig{figure=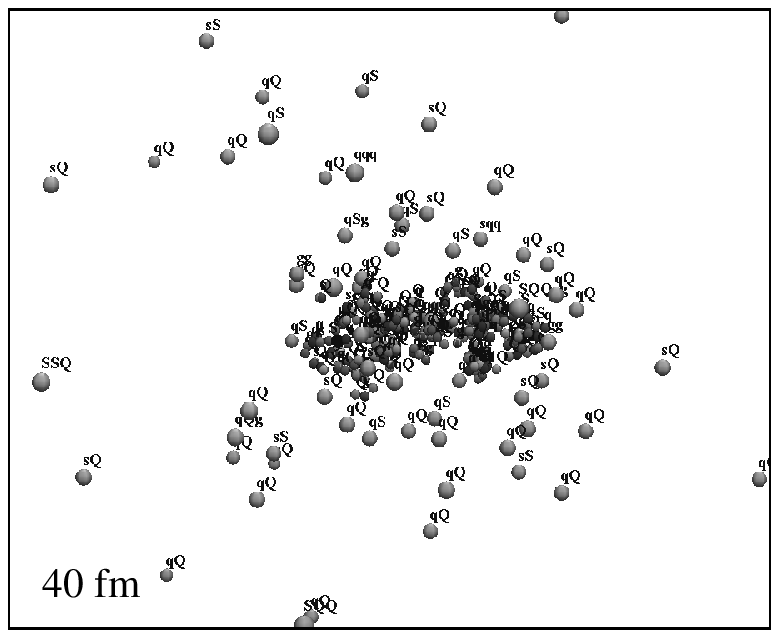,width=80mm}
            \psfig{figure=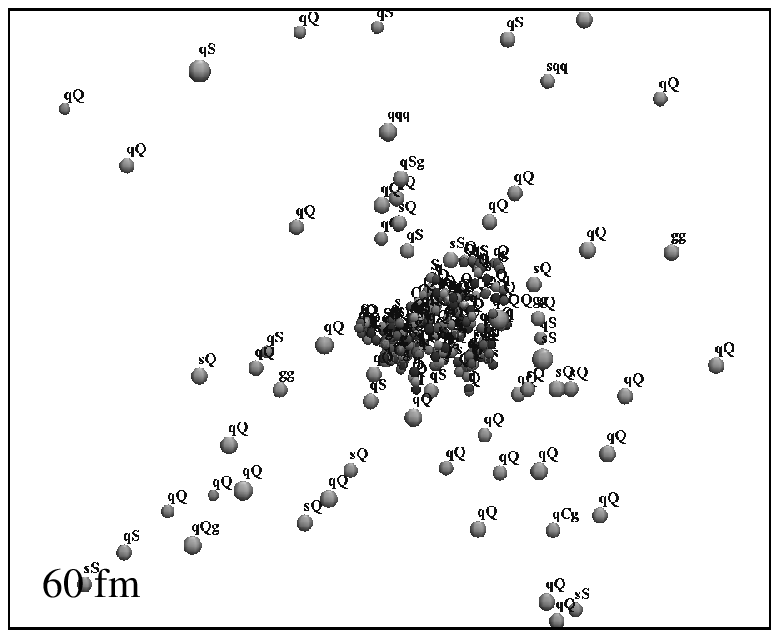,width=80mm}}
\centerline{\psfig{figure=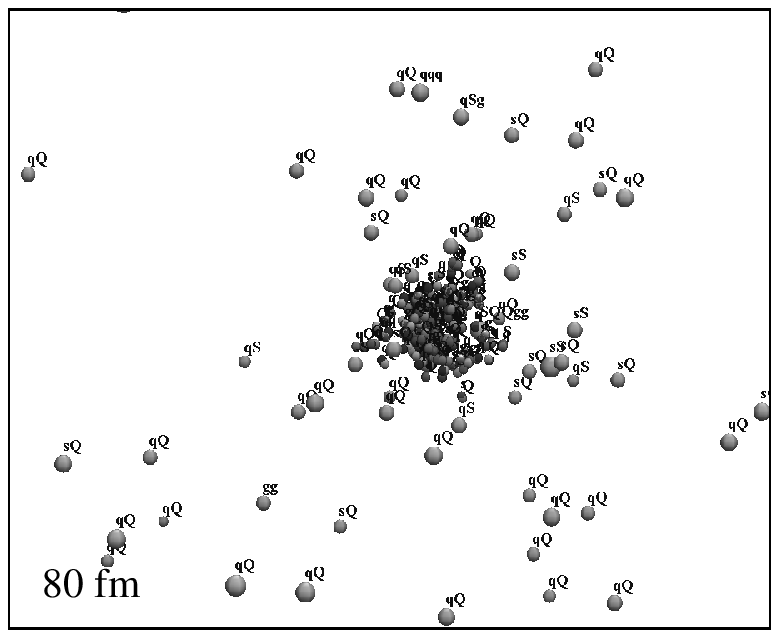,width=80mm}
            \psfig{figure=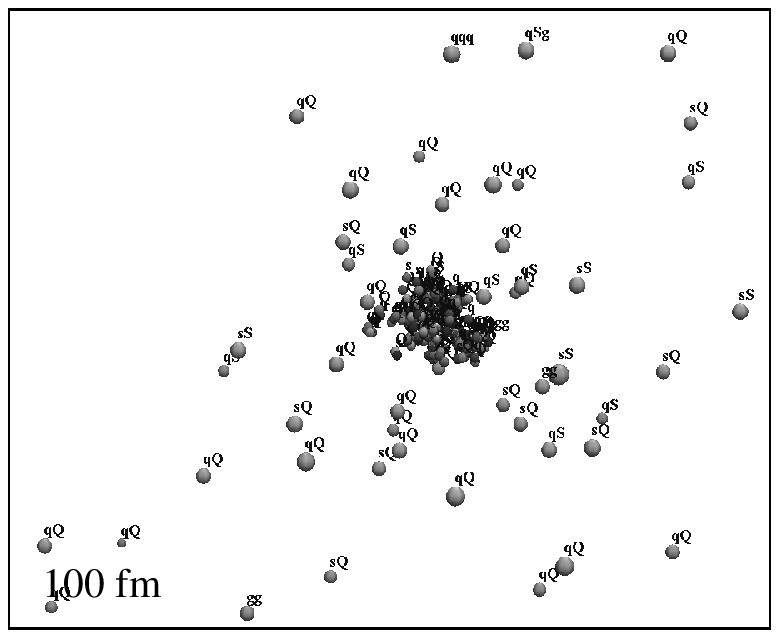,width=80mm}}
\caption{\em Simulation snapshots after initialization of
a central-rapidity slice of a QGP tube with Lorentz-invariant flow
(``Bjorken scenario''). The frame diagonal is of length 45 fm.
The labels at the particles and hadrons
indicate the various flavors and constituents; the letter $q$
represents an up or down quark; the other letters have obvious
meanings, capital letters denoting antiquarks.
Colored objects (quarks, gluons) are drawn in dark grey,
hadrons in light grey.}
\label{bjorkensnapshots}
\twocolumn
\end{figure}

\begin{figure}
\onecolumn
\centerline{\psfig{figure=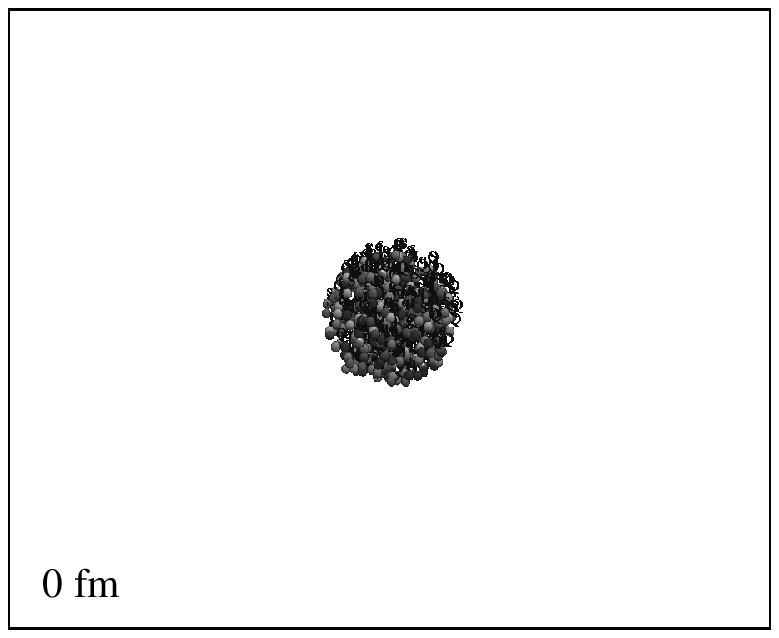,width=80mm}
            \psfig{figure=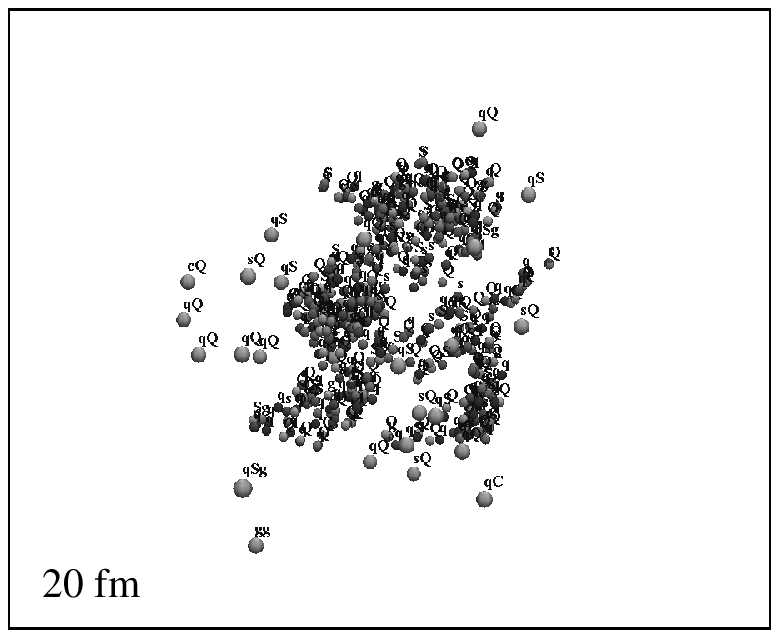,width=80mm}}
\centerline{\psfig{figure=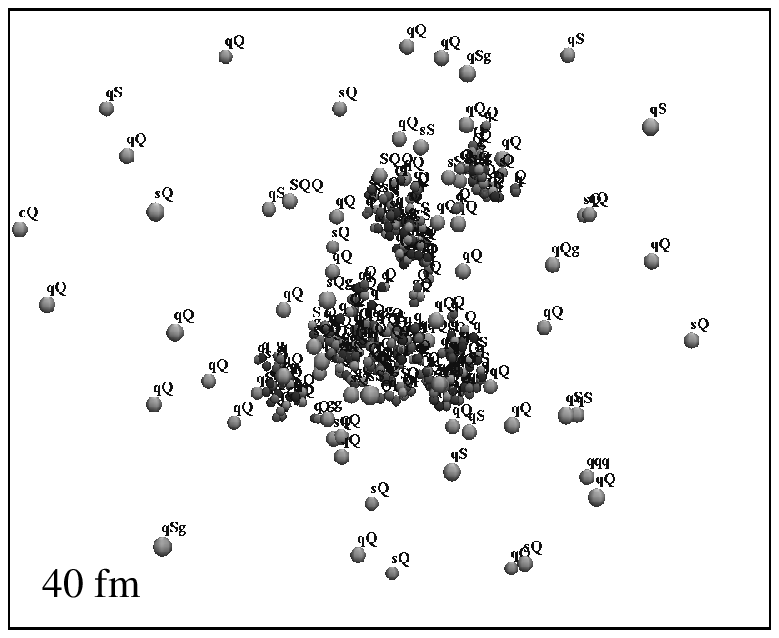,width=80mm}
            \psfig{figure=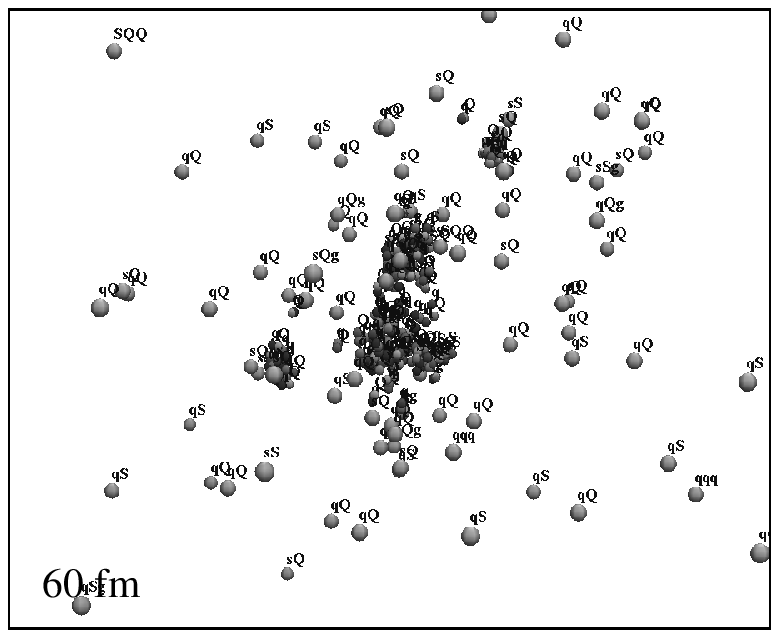,width=80mm}}
\centerline{\psfig{figure=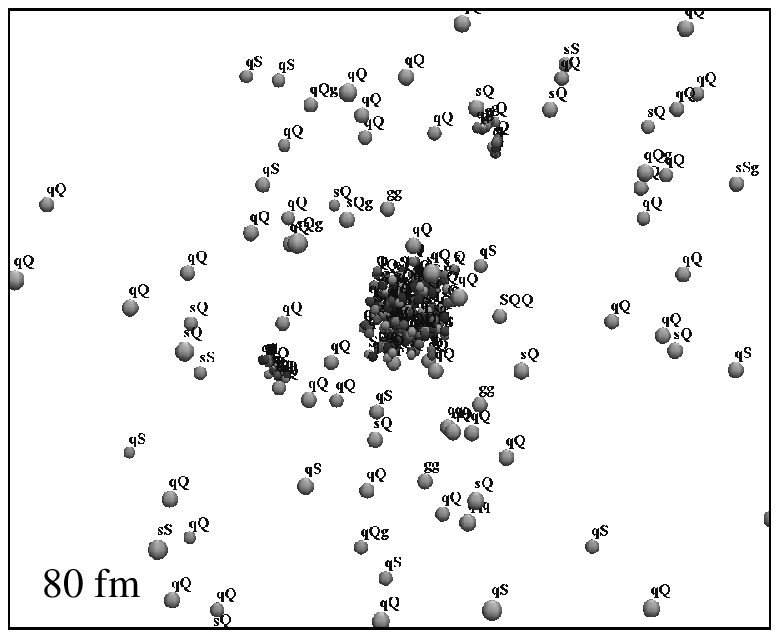,width=80mm}
            \psfig{figure=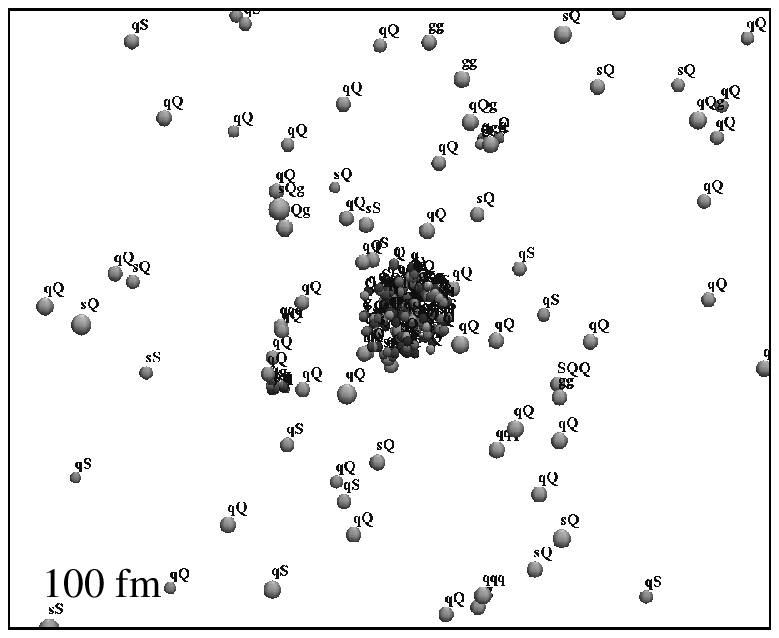,width=80mm}}
\caption{\em Simulation snapshots after initialization of
a thermal QGP fireball (``full stopping scenario'').}
\label{stoppingsnapshots}
\twocolumn
\end{figure}

\vvs
\subsection{Example Simulation Snapshots}

Figs.~\ref{bjorkensnapshots} and \ref{stoppingsnapshots} 
show six snapshots of the time evolution for
the Bjorken and ``full stopping'' scenario\footnote{By the time
this article will be printed, MPEG movies of the actual simulation 
runs will be retrievable by anonymous ftp from
{\tt ftp://}{\tt theorie}{\tt.physik}{\tt.uni-giessen}{\tt.de}
{\tt/pub}{\tt/cdmmovie}{\tt/bjorken.mpg} resp. {\tt/stopping.mpg}}. 
The simulation times of the snapshots are written inside the frames 
in units of fm; the frame diagonal is of length 45 fm.
Quarks, gluons, and hadrons are drawn as
spheres; for better visibility of the scene, the radius of a colored
object is scaled down to one third of the actual RMS radius of the charge
distribution (0.7 fm). Hadron radii reflect the bag volume formed by 
the $\sigma$ field footprint of the hadronized cluster; they are
scaled down to one half of their actual size. For white
clusters without any $\sigma$ field bag (see section
\ref{IrreducibleWhiteClusters}), we draw a sphere of radius 0.35 fm.
The $\sigma$ field as well as the two color 
fields are not displayed. Note that in the last frame, many hadrons
have already left the scene.

As can be seen from the pictures, the initial partonic fireball is
expanding and hadronizing. At each timestep, the equations of motion of the charges
and the $\sigma$ field as well as the Gauss law is solved
(see Sec.~\ref{NumericalSolutionOfFieldEquations}). Whenever an
irreducible white cluster is detected (by the mechanism described in
Sec.~\ref{IrreducibleWhiteClusters}), it is classified according to
Fig.~\ref{clusters} as a meson, baryon or glueball, its total energy, momentum
and mass are computed, the partons and their field imprint are
removed from the simulation and replaced by a noninteracting
``hadron'', displayed in the snapshots as a grey sphere.
We would like to emphasize that the production of hadronic states is
not artificially put in by hand: the model itself produces irreducible
white clusters in the final state, and it is only up to our detection
routine to classify them and take them out of the simulation.
The detection routine does not drive or interfere with the process of hadron
formation, it merely serves to interpret the results.
Although the last snapshot still shows quarks and gluons,
this is not the final state of the simulation run. In the final state
of the simulation run, hadronization is indeed complete: no partonic 
fireball is left over.

\vvs
\subsection{Energy Conservation, Hadronization Time, and Temperature}
\label{EnergyConservation}

\begin{figure}
\centerline{\psfig{figure=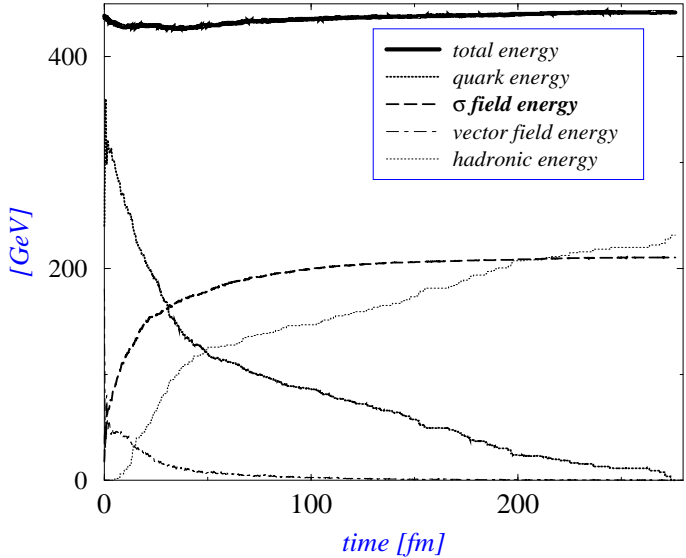,width=90mm}}
\caption{\em System energy during time evolution in the Bjorken event.}
\label{energy}
\end{figure}

\begin{figure}
\centerline{\psfig{figure=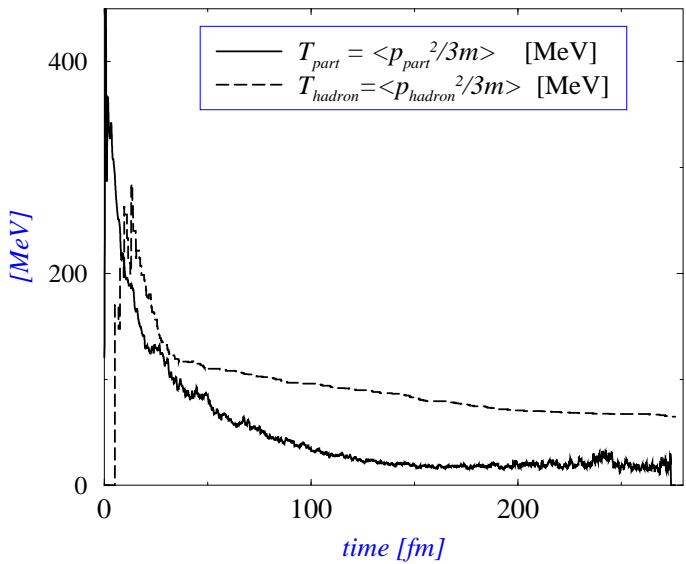,width=90mm}}
\caption{\em Particle and hadron temperature over time in the Bjorken event.}
\label{temperature}
\end{figure}

Fig.~\ref{energy} shows the energy over time of the system and its
parts in the Bjorken scenario, for the same event the snapshots were
taken from.
Fig.~\ref{temperature} shows for the same time interval
the temperature evolution
of the partons in the fireball and the produced hadrons.
Their temperature is found from
\be
{3\over2}T\ :=\ \left\langle{{\vec p\,}^2\over2m}\right\rangle.
\label{KineticTemperature}
\ee

The partons heat up initially because
they are accelerated by relatively long ($>$ 1 fm) flux tubes. These
flux tubes are present simply because of the initial spatial random
distribution of the charges. Very soon, however, the hot particles
cool down, transferring energy to the
$\sigma$ field. On the same time scale, the energy contribution
of the color electric field, also shown in Fig.~\ref{energy},
decreases to almost zero. This is an indication
of the strong electric shielding of the plasma. The picture of the plasma
as a collection of flux tubes breaks down here; flux tubes emerge only
on the surface of the fireball, pulling back colored objects that
try to escape. It is interesting to note that, although the
energy contribution of the color field itself is never
significant, the color field dictates the dynamics of the
hadronization process, allowing only white objects to separate.

The hadronic energy contribution is a monotonous step
function: it increases each
time we take another quark-gluon cluster, delete its field imprint and
add it to the set of hadrons in the system. The hadrons do no longer
interact, hence the piecewise constant energy.
About half of the final energy is deposited in $\sigma$ waves. This
energy has its origin in the initialized thermal energy of the color
charges. These form flux tubes, and via the coupling term
$\kappa(\sigma)F^2/4$ the flux tube cools down, radiating $\sigma$
waves. The $\sigma-$field describes the long-range collective effects
of QCD. Its large energy content shows that the nonperturbative QCD
vacuum is highly excited; in the ``real world'', it would decay mostly
into pions which are so far not contained in our simulation.

Note that most of the final hadronic energy is produced in the
time interval $[10{\rm fm}, 50 {\rm fm}]$. These 40 fm could be interpreted as
the hadronization time in our model.

There is no final-state interaction among the
hadrons or between hadrons and partons
in the simulation, so the hadronic subsystem is not in thermal
contact with the fireball. This is realistic: the hadron temperature
reflects the parton
temperature at the time of hadron formation (freeze-out point).
Since most of the hadrons
are produced in the time interval $[10 {\rm fm}, 50 {\rm fm}]$, the final hadronic
temperature reflects the fireball temperature at that time. The
fireball cools further down, which leads to the unrealistic picture
that after about 50 fm, the fireball is significantly cooler than the
hadrons.

In an experiment, a fireball as cold as $50 MeV$ would
probably hadronize immediately. In our simulation, however,
hadrons are only well-defined once they have clearly left the fireball
and can be {\em detected} as a
separate, irreducible white cluster.
At this late stage of the simulation run, hadronization is a random
evaporation process:
inside the fireball, color fields are strongly shielded,
which prevents stable white clusters to form.
So an irreducible white cluster (hadron) can only form at the surface
of the fireball, as soon as its constituents, floating around
in the fireball nearly freely,
happen to reach the surface with coalescent momenta.
Thus it takes quite some
time for a small, cold parton cluster to hadronize completely.

We attribute this partly to the missing repulive interaction between
the white clusters in our simulation \cite{KOEPF}. In addition, it is
well known from nuclear fragmentation studies that the mean field
tends to prevent clusterization \cite{JUNG}. Thus, fluctuations of the
color fields may be important for the final hadronization process.

%We feel that further tweaking of the model parameters won't
%change this outcome drastically.
%This raises the question: which crucial feature that is missing from our model
%causes nature to hadronize so much faster?
%With our present understanding of the model, we can only exclude one
%possible answer:
%A particle production term, certainly present in nature and
%missing from our simulation, will not speed up hadronization.
%On one hand,
%production of two opposite charges (of appropriate colors)
%can break a flux tube.
%Thus a single parton, say a quark, of high momentum can leave the
%fireball, if the flux tube (that tries to pull the quark back) is
%broken apart by $q\bar q$ production. The antiquark then forms a meson
%together with the initial quark.
%However, this process would not speed up hadronization of the fireball,
%because the newly produced quark would afterwards belong to the
%fireball. So although a fast parton can quickly hadronize by flux tube
%breaking, the total number of particles in the fireball is not reduced
%by such a process. We conclude that a particle production term
%can speed up cooling but not hadronization of the fireball.

\vvs
\subsection{Mass distribution}

\begin{figure}
\centerline{\psfig{figure=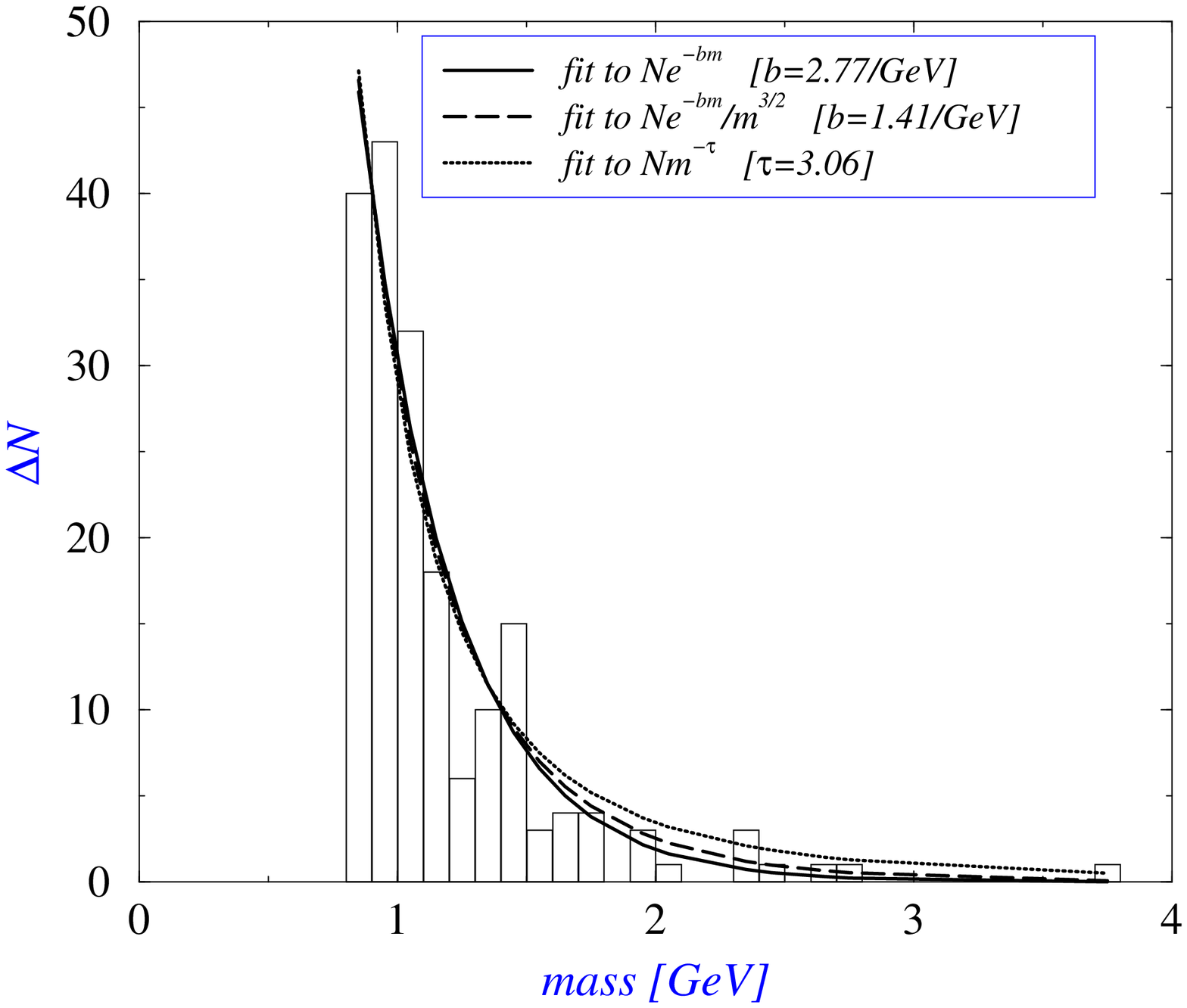,width=90mm}}
\centerline{\psfig{figure=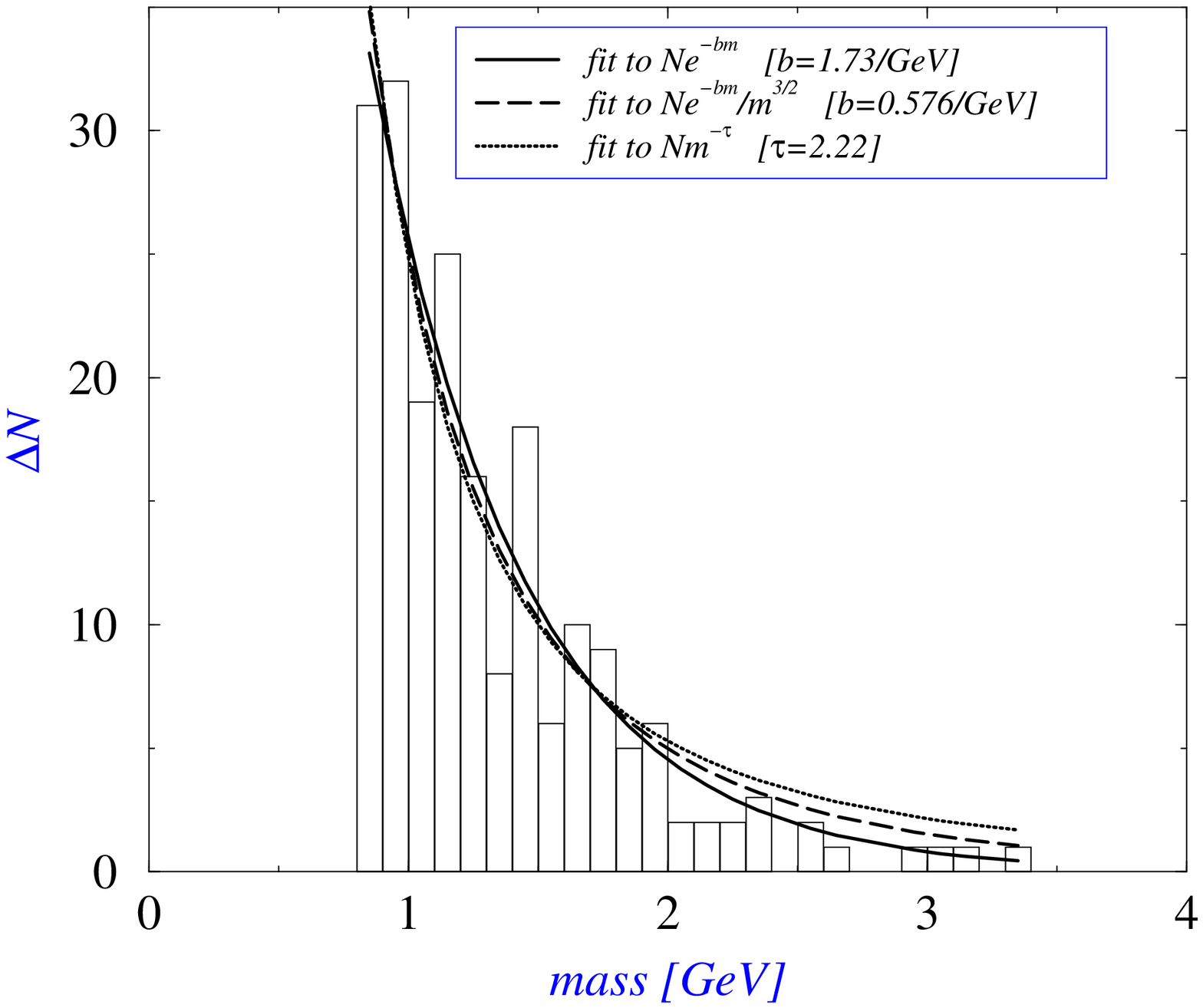,width=90mm}}
\caption{\em Mass spectra; the upper graph is the Bjorken case, the
lower the ``full stopping'' scenario.}
\label{mass_spectra}
\end{figure}

The final mass distribution of the hadrons is shown in
Fig.~\ref{mass_spectra} for both the Bjorken and the ``full stopping''
scenario. The spectra start at a mass of $800$ MeV because this is
the mass of the lightest hadron ($q\bar q$) in our model, and no
quarks or gluons are left over in the final state.

Although the spectra do look different on first sight,
one should keep in mind that since these are single events, the
differences in the spectra may simply be due to statistical 
fluctuations and might not be significant of the respective scenario.
Note also that the shape of the mass spectrum in the Bjorken 
case would stay the same if the chosen rapidity slice
$\Delta y=[-0.5, 0.5]$ were extended to a realistic interval,
simply because masses are Lorentz invariant.

Assuming a Hagedorn density of hadronic states \cite{HAGEDORN},
\be
 \rho(m)\ \propto\ m^a e^{m/T_H},
\label{HagedornDensityOfStates}
\ee
we expect in the nonrelativistic limit a measured mass distribution 
\ba
 {dN\over dm}&\propto&\rho(m)\int d^3p\, e^{-(m+p^2/2m)/T} \nonumber \\
  &\propto& m^{a+{3\over2}} e^{-m\left({1\over T}-{1\over T_H}\right)}.
\label{HagedornMassSpectrum}
\ea
The nonrelativistic limit is justified by our large constituent masses.
Here, $T_H$ is the Hagedorn temperature: the limit temperature of
hadronic strings. Beyond this temperature, the flux tube picture must
be replaced by a true QGP. We assume $T_H=160$ MeV for now 
(which is what Hagedorn estimated originally) and take
$a=-3$ from \cite{GORENSTEIN}. Fig.~\ref{mass_spectra} shows that a
satisfactory fit of the form (\ref{HagedornMassSpectrum}) can be
obtained. Our fit parameter $b$ is related to $T$ via 
\be
    b\ =\ {1\over T}-{1\over T_H}.
\ee
The corresponding hadron temperature is $T=130$ MeV for the Bjorken
event ($T=146$ MeV for the full stopping event), which is
somewhat higher than the kinetic temperature found in the last
section. However, these numbers depend on the choice of $a$:
taking $a=-3/2$, we expect a simple exponential curve for the mass
spectrum, which results in different values for $T$, namely $T=110$
MeV for the Bjorken event and $125$ MeV for the full stopping
event. In this single-event analysis with our very small statistics,
we cannot really decide on the correct value of $a$.

Another possible expectation for the mass spectrum could be that of
a percolative process near the critical point \cite{STAUFFER}. There, a 
scaling distribution
\be
 {dN\over dm}\ \propto\ m^{-\tau}
\label{PercolationMassSpectrum}
\ee
is expected. Trying to fit this relation to our data (see
Fig.~\ref{mass_spectra}), we find a satisfactory agreement for
$\tau=3.06$ (Bjorken case) and $\tau=2.22$ (full stopping event). 
We plan to investigate this further in the future.

\vvs
\subsection{Rapidity and Momentum Spectra}

\begin{figure}
\centerline{\psfig{figure=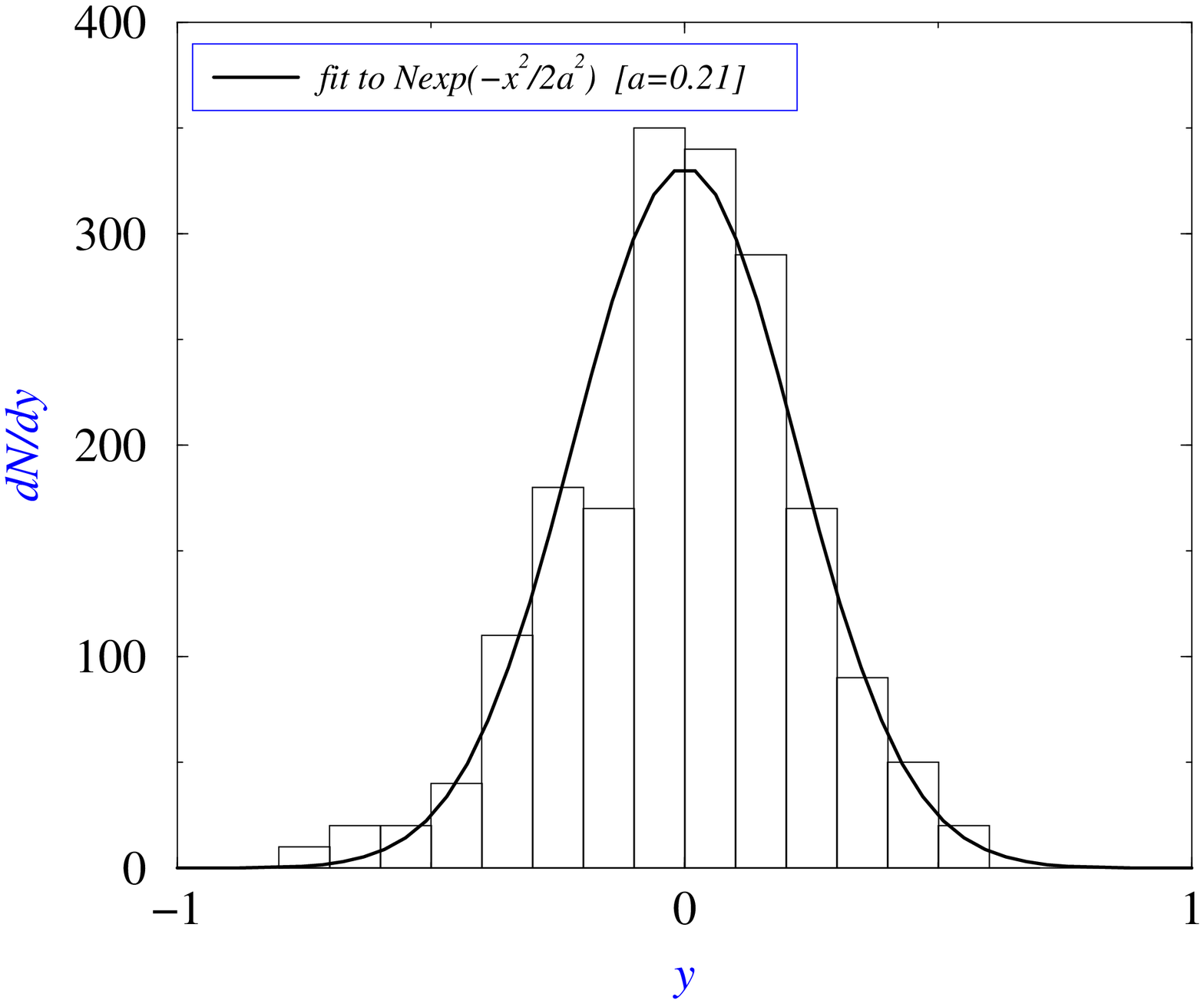,width=90mm}}
\centerline{\psfig{figure=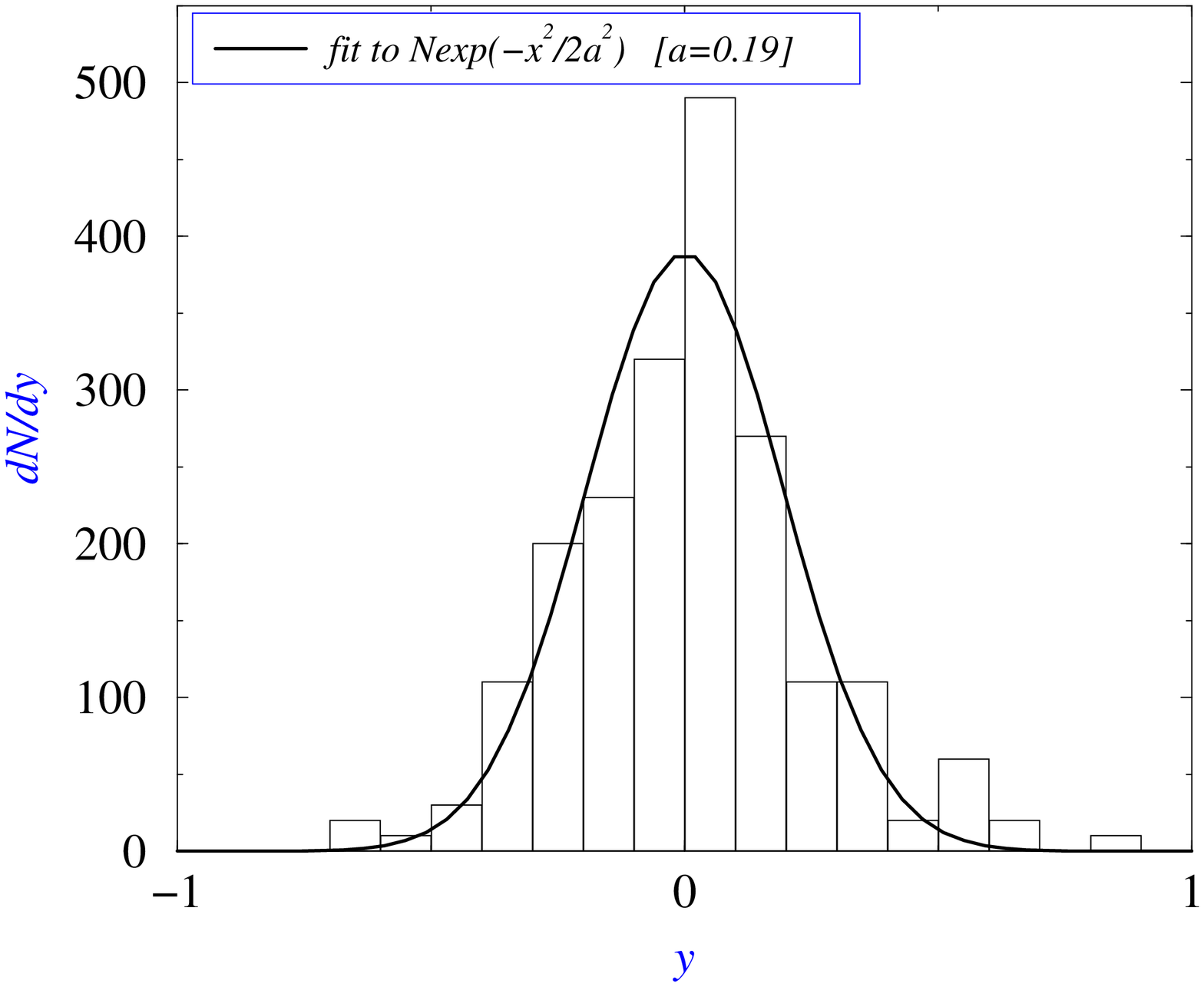,width=90mm}}
\caption{\em Rapidity spectra; the upper graph is the Bjorken 
case, the lower the ``full stopping'' scenario.}
\label{rapidity_spectra}
\end{figure}

Fig.~\ref{rapidity_spectra} shows the hadron rapidity spectra of 
both scenarios. They do not look significantly different, the Bjorken
case being somewhat broader due to the longitudinal flow.
However, no plateau structure is visible in the Bjorken
case, since we have cut out a central-rapidity slice of the Bjorken
tube. One could easily extrapolate this result by folding the
rapidity spectrum with that of a Bjorken QGP tube of 
realistic extension. This folding process would then result in a plateau 
structure.

\begin{figure}
\centerline{\psfig{figure=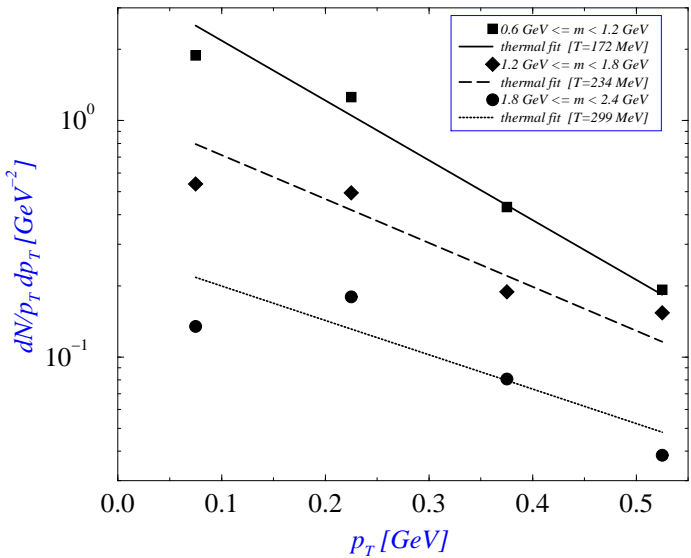,width=90mm}}
\caption{\em Final transversal momentum spectrum for the Bjorken case,
for three mass bins separately. When computing the fit lines, we took
the statistical error bars correctly into account \protect\cite{NUMREC}.}
\label{pt_spectra}
\end{figure}

\begin{figure}
\centerline{\psfig{figure=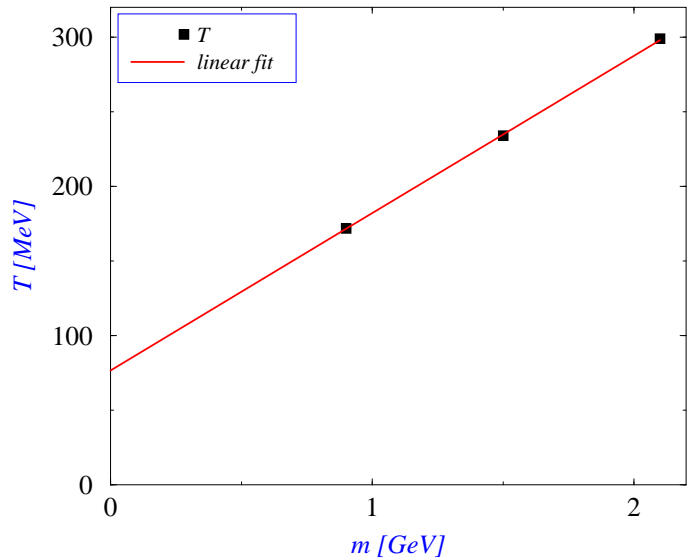,width=90mm}}
\caption{\em The temperatures extracted from mass differential $p_T$ spectra,
plotted against hadron rest mass.}
\label{sorgefit}
\end{figure}

Finally, Fig.~\ref{pt_spectra} shows the transverse momentum spectra of 
nonstrange white hadronic clusters in the Bjorken scenario, 
observed in mass bins of width $0.6$ GeV above $0.6$ GeV.
Although the points in the Figure representing  mid-$p_T$ values do not lie
exactly on exponential curves,  it is interesting to consider
the slope values in order to extract possible information
on the flow and temperature of the fireball in the late
hadronization phase.
The resulting inverse slopes, denoted as ``temperatures'' $T(m)$,
are displayed in Fig.~\ref{sorgefit}. By fitting to the linear relation
\be
  T(m) = T_0 + m\langle v_T^2\rangle,   
\ee
we extract from the plot a temperature $T_0=77$ MeV and a radial flow 
$\sqrt{\langle v_T^2\rangle}=0.32$. 
A direct computation of the radial flow from our final state yields
$\sqrt{\langle v_T^2\rangle}=0.30$, which is remarkably close.
For the full stopping event, the same $T(m)$ analysis yields 
$T_0=126$ MeV and $\sqrt{\langle v_T^2\rangle}=0.18$, 
while the true radial flow is $\sqrt{\langle v_T^2\rangle}=0.27$, a
not so good agreement. 
Compared to data on SPS heavy ion experiments at CERN \cite{CERNFLOW}
($\langle v_T \rangle = 0.4 $, $v_{{\rm max}} = 0.6$ and $T = 140$ MeV)
our results display a little less flow and a colder hadronic fireball.
It should be noted, however, that this analysis is based on a single
event whose plasma volume
is only a fraction of the experimental one.
Also, we extracted the spectra of rather heavy hadronic clusters;
their decay products (with lower mass) may have higher
$p_T$-slopes leading altogether to a higher estimate of $T_0$.
From this viewpoint, the qualitative agreement is satisfactory.

%%%%%%%%%%%%%%%% CONCLUSION %%%%%%%%%%%%%%%%
\vvs
\section{Conclusion}
\label{Conclusion}

\vs
We have investigated the hadronization process of a quark-gluon
droplet in the framework of the chromodielectric model, 
using a classical molecular dynamic approach.
An efficient computer code was written and put to work
for the solution of the boundary value problem connected to the Gauss law
in the abelian dominance approximation.
This code made it possible to study a volume
of about $(25 {\rm fm})^3$, filled with about 400 particles.

Our quark masses were fitted to a constituent model
of the vector meson octet and baryon decuplet states.
We determined from the microscopic data of hadronization
events the mass-, rapidity- and transverse momentum distributions.
  
As example scenarios we have taken two different initial states,
one with full stopping and one with a one-dimensional
scaling flow. The two scenarios behave similarly, having
a commensurate energy density initially.
We found that although the identification of white clusters
technically requires a long simulation time, the mass
distribution develops a hadron part already at the
beginning of the process --- about 10 fm after initialization
--- and the main hadronization process is done after another
25 fm have passed.
In a recent RQMD paper \cite{SORGE} similar hadronization
times are reported.

Hadronization is certainly a much faster process than this
simulation suggests. This is because we can only
uniquely identify an irreducible white cluster after it has already
left the fireball. So in this simulation, hadronization looks like a
slow evaporation effect, while in principle pre-hadrons could already
form in the plasma but remain undetected by our program.
%As argued in Sec.~\ref{EnergyConservation}, we do not expect
%a future particle production term to significantly change this picture.
An effective, quantum-mechanical, repulsive force between white 
clusters, as computed by Koepf et al. \cite{KOEPF}, could change this 
picture, if a consistent way of including this in our simulation can
be found.
	
Summarizing, we do not observe a vast
departure from thermal equilibrium during this
microscopic study of hadronization. Chemical equilibration
on the other hand is incomplete as shown by the deviation
between the Maxwell-Boltzmann slopes for different cluster
species and in general between colored and white particles.
This also supports the conclusion that during the
late hadronization of quark matter there is no
phase equilibrium between the quark-gluon and hadron components;
the Gibbs criteria do not apply. The suggested picture
is rather a mixture of hadronic and quark-gluon plasma
counterparts (with additive instead of equal pressure).

An extended study of cluster
distribution in the near future is planned in
order to investigate whether the hadronization is
a percolation- or phase-transition-like process.

%%%%%%%%%%%%%%%%%%%%% THANX %%%%%%%%%%%%%%%%%%%%%%%%%%%%%%

\section{Acknowledgments}

This work is part of a collaboration between the
Deutsche Forschungsgemeinschaft and the Hungarian Academy of
Science (project No. DFG-MTA 101/1998) and has been
supported by the Hungarian National Fund for Scientific
Research OTKA (project No. T019700 and T024094) as well as the
Bundesministerium f\"ur Bildung, Wissenschaft, Forschung und
Technologie (BMBF) and the Gesellschaft f\"ur Schwerionenforschung
(GSI) Darmstadt.

Discussions with C.~Greiner, S.~Leupold and J.~Zim\'anyi are
gratefully acknowledged.

%%%%%%%%%%%%%%%%%%% REFS %%%%%%%%%%%%%%%%%%%%%%%%%%%%%%%

\end{document}